\documentclass{article}
\PassOptionsToPackage{numbers, compress}{natbib}

\usepackage[preprint]{neurips_2025}

\usepackage[utf8]{inputenc} 
\usepackage[T1]{fontenc}    
\usepackage{hyperref}       
\usepackage{url}            
\usepackage{booktabs}       
\usepackage{amsfonts}       
\usepackage{nicefrac}       
\usepackage{microtype}      
\usepackage{xcolor}         
\usepackage{enumitem}
\usepackage{graphicx}

\usepackage{colortbl}  
\usepackage{xcolor}  
\usepackage{amsmath}
\usepackage[utf8]{inputenc}
\usepackage[most]{tcolorbox}
\usepackage{pgfplots}
\usepackage{enumitem}
\usepackage[table]{xcolor}
\usepackage{latexsym}
\usepackage{wrapfig,lipsum}
\usepackage{pifont}
\usepackage{booktabs}
\usepackage{amssymb}
\usepackage{times}
\usepackage{adjustbox}
\usepackage{multirow}
\usepackage{caption}
\usepackage{subcaption}
\usepackage{float}
\definecolor{nBlue}{RGB}{0,165,249}
\definecolor{nGreen}{rgb}{0, 0.5, 0.2}
\definecolor{nRed}{rgb}{0.8, 0.1, 0.2}
\definecolor{mGreen}{rgb}{0.3, 0.65, 0.4}
\usepackage[ruled, vlined, noend, linesnumbered]{algorithm2e} 
\SetKwFor{Upon}{upon}{do}{end upon}
\usepackage{bbm}

\title{\textsc{Ecdysis}: Efficient and Effective Training of Runtime Harnesses for LLM Agents}

\author{ 
\textbf{Ruiqing Yue\textsuperscript{1,2}}\thanks{Yu Cui and Ruiqing Yue are co-first authors and contributed equally to this work. Yu Cui proposed the algorithm, and Ruiqing Yue validated its effectiveness through experiments.} \quad 
\textbf{Yu Cui\textsuperscript{3}}\footnotemark[1] \quad 
\textbf{Zhuoyu Sun\textsuperscript{3}} \quad 
\textbf{Sicheng Pan\textsuperscript{3}} \quad
\textbf{Xianhong Xue\textsuperscript{1,2}} \quad \\
\textbf{Tingyu Li\textsuperscript{3}} \quad 
\textbf{Ting Li\textsuperscript{4}} \quad 
\textbf{Wenzhuo Zhu\textsuperscript{4}} \quad 
\textbf{Yi Chen\textsuperscript{1,2}} \quad 
\textbf{Yifei Liu\textsuperscript{3}} \quad  \\
\textbf{Baohan Huang\textsuperscript{3}} \quad 
\textbf{Zhe Cui\textsuperscript{1,2}} \quad 
\textbf{Haibin Zhang\textsuperscript{5,6}} 
\quad \textbf{Cong Zuo\textsuperscript{3}}\quad \\ 
\small\textsuperscript{1}Chengdu Institute of Computer Applications, Chinese Academy of Sciences\\ \small\textsuperscript{2}University of Chinese Academy of Sciences 
\quad \small\textsuperscript{3}Beijing Institute of Technology \\ 
\small\textsuperscript{4}Beijing University of Technology \quad 
\small\textsuperscript{5}Yangtze Delta Region Institute of Tsinghua University, Zhejiang \\ 
\small\textsuperscript{6}Jiaxing Key Laboratory of Artificial Intelligence and Cyber Resilience\\ 
\url{https://github.com/cuiyu-ai/Ecdysis}\\ 
\small{\textbf{Project Lead}: Yu Cui <cuiyu@bit.edu.cn>} 
}

\begin{document}

\maketitle

\begin{abstract}
Self-evolving runtime harnesses can substantially improve the capabilities of large language model (LLM) agents and provide a promising paradigm for optimizing agent execution. Existing failure-driven approaches often treat observed agent failures as direct evidence for harness modification. A key challenge in failure-driven harness evolution is that observed failures can reflect either limitations of the underlying model or systematic deficiencies of the harness. Directly optimizing against individual failures can therefore induce model-specific accommodation and impair generalization across tasks and models. We study whether failure evidence accumulated across task instances can provide a more reliable signal for harness training. Our key insight is that failures recurring across distinct tasks provide stronger inductive evidence for systematic harness deficiencies than isolated failures. Based on this insight, we propose \textsc{Ecdysis}, which aggregates failure evidence across task instances before promoting recurring failure patterns into persistent harness evolution, biasing evolution toward repairs that are more likely to generalize beyond individual model behaviors. \textsc{Ecdysis} further employs collaborative failure analysis to refine modification specifications, trading additional evolution-time reasoning for improved modification quality. Across multiple LLMs and benchmarks, \textsc{Ecdysis} improves the reasoning accuracy of evolved harnesses by 18.56\% over existing harness evolution while achieving up to 1.84$\times$ faster harness training. \textsc{Ecdysis} also enables more data-efficient training. Fine-grained analysis shows that \textsc{Ecdysis} reduces model-specific accommodation during evolution, while the resulting harnesses exhibit stronger cross-LLM generalization and lower inference-time token consumption. These results suggest that effective harness training depends not only on how failures are repaired, but also on which failures provide reliable evidence for persistent harness changes.
\end{abstract}

\section{Introduction}

Large language models (LLMs) have demonstrated increasingly strong capabilities in complex reasoning, yet practical LLM agents often rely critically on runtime harnesses that govern task planning, tool interaction, and context management \citep{chen-etal-2025-locagent, jiang2026self, yao2023react, wan-etal-2026-compass}. Recent harness self-evolution methods \citep{lee2026recursive, chen2026failed, wang2026huxleygodel, shao2026harness} enable these runtime mechanisms to improve automatically from execution feedback, without modifying the underlying model. By iteratively identifying failures and revising the harness, this paradigm provides a promising approach to optimizing agent behavior. However, failure-driven evolution also introduces a fundamental challenge: execution failures must be correctly interpreted before being translated into persistent harness modifications. Otherwise, repeated adaptation to observed failures may lead the harness to become specialized to particular tasks or model behaviors, increasing evolution cost and potentially impairing generalization \citep{shao2026your, jiang2026harnessevolve, wang2026rethinking}.

More fundamentally, \textit{an observed agent failure does not necessarily indicate a harness deficiency}. A failure may instead reflect a limitation of the underlying model. Modifying the harness to accommodate such behavior may improve performance on the current model while unnecessarily coupling the harness to it, potentially harming generalization to unseen tasks and other LLMs. We distinguish between \emph{model-specific accommodation} (MSA), which compensates for model-level limitations, and \emph{harness-level repair} (HLR), which addresses deficiencies in the runtime mechanism \citep{zhang2026harnesscompass}. Individual failures provide limited evidence for distinguishing between these two cases. In contrast, failures that recur across distinct task instances provide stronger inductive evidence that the underlying mechanism, rather than an isolated model behavior, may be responsible. This observation suggests that persistent harness evolution should be driven not simply by whether a failure occurs, but by whether the failure constitutes sufficiently general evidence for a harness-level deficiency.

This motivates our research question: \textbf{Can cross-task failure evidence guide harness training toward systematic repair while reducing unnecessary MSA?} We propose \textsc{Ecdysis}, a framework that shifts failure-driven harness training from instance-level correction toward cross-task evidence aggregation. \textsc{Ecdysis} jointly analyzes failures across a batch of task instances, identifies recurring failure patterns, and uses them to prioritize candidate harness modifications. \textsc{Ecdysis} biases training toward changes supported by repeated failure mechanisms rather than isolated model behaviors. It further employs collaborative failure \citep{yu-etal-2025-table, chen-etal-2025-magicore, zhang-etal-2025-enhancing-recommendation} analysis to refine modification specifications. This refinement is complementary to the central aggregation mechanism: \textsc{Ecdysis} first determines \emph{which failure evidence should drive evolution}, and then \emph{improves how that evidence is translated into harness modifications}.

We systematically evaluate \textsc{Ecdysis} across multiple LLMs and benchmarks. Experimental results show that \textsc{Ecdysis} achieves up to a 1.84$\times$ speedup in harness training compared with existing harness evolution methods while improving the reasoning accuracy of the resulting harnesses by 18.56\%. Further experiments demonstrate that harnesses trained with \textsc{Ecdysis} exhibit stronger cross-LLM generalization, while also reducing inference-time token consumption. The cross-task perspective also enables failure-aware training data curation. This indicates that the informativeness and recurrence structure of failure evidence can matter more than the amount of failure data alone. 
Notably, the ablation study reveals a clear trade-off between efficient evidence aggregation and reliable failure attribution. With collaborative refinement, MSA decreases from 75.61\% to 45.45\%, leading to stronger cross-LLM generalization of the evolved harnesses, at the cost of additional evolution-time computation. These results indicate that recurrence provides an efficient mechanism for selecting failure evidence, but recurrence alone is insufficient for reliably distinguishing recurring model limitations from harness deficiencies. Explicit failure attribution through collaborative refinement is therefore complementary to cross-task aggregation. Our contributions are summarized as follows:

\begin{itemize}[left=0pt, itemsep=0pt]

\item \textbf{Cross-Task Failure Recurrence for Efficient Evidence Selection}.
We propose cross-task failure recurrence as an efficient mechanism for selecting failure evidence that warrants persistent harness evolution. By aggregating failures across task instances, \textsc{Ecdysis} substantially reduces the evolution cost and training time required to identify recurring failure patterns. However, recurrence alone may over-amplify recurring model-specific failures, revealing the need for explicit failure attribution before translating recurring evidence into persistent harness modifications.

\item \textbf{Collaborative Refinement for Failure Attribution}.
We introduce a collaborative refinement that analyzes recurring failure evidence before implementation, providing an explicit stage for distinguishing how failures should be interpreted and translated into harness modifications. Our experiments show that this additional reasoning incurs computational overhead but improves final harness accuracy and cross-LLM generalization while substantially reducing MSA.

\item \textbf{Failure-Aware Data Curation}.
We show that recurrence-aware failure analysis can guide training-data selection, allowing effective harness evolution with substantially fewer failure instances.

\end{itemize}

\section{Related Work}
\label{sec:related-work}
Inference-time mechanisms such as tool interaction, deliberate planning, and feedback-driven adaptation have substantially enhanced LLM agents \citep{yao2023react, shinn2023ref}. Building on these advances, automatic optimization has progressed from prompt optimization \citep{khattab2024dspy,yang2024large} to agentic workflows and self-modifying agent programs \citep{hu2025adas,zhang2025aflow,zhang2026darwin}. More recently, several works have focused specifically on evolving runtime harnesses, using execution feedback and iterative evaluation to propose and validate harness modifications \citep{zhang2026self, lee2026recursive, chen2026harnessforge, wang2026handbook, huang2026memo, xu2026verify}. 
Although harnesses are ultimately optimized to improve LLM task performance, harness evolution should not simply adapt to every observed failure, as failures may reflect model limitations rather than harness deficiencies, leading to MSA. \textsc{Ecdysis} therefore treats recurrent failure patterns across independent task instances as stronger evidence of systematic harness deficiencies, guiding evolution toward reusable behavioral improvements rather than instance-specific adaptation. Self-Harness \citep{zhang2026self} treats model-specific weaknesses as actionable targets for harness adaptation, whereas \textsc{Ecdysis} seeks to distinguish such accommodation from persistent harness deficiencies. \textsc{Ecdysis} differs from HarnessCompass \citep{zhang2026harnesscompass} in how it controls over-specialization during harness evolution. HarnessCompass explicitly constrains the evolution search space to rule out task-specific modifications, whereas \textsc{Ecdysis} introduces no such hard constraint. Instead, \textsc{Ecdysis} biases the evolution process toward systematic harness deficiencies by aggregating failure evidence across distinct task instances and retaining recurring failure patterns as actionable signals. Consequently, MSA is mitigated rather than prohibited: \textsc{Ecdysis} reduces the influence of idiosyncratic model-specific failures on persistent harness evolution while preserving potentially useful adaptations when supported by recurring evidence.

\section{\textsc{Ecdysis}}
\definecolor{nGreen}{RGB}{0,120,70}
\newcommand{\InlineComment}[1]{\textnormal{\hspace{0.45em}\color{nGreen}\#~#1}}

\subsection{Preliminary Analysis}
\label{sec:preliminary-analysis}
A fundamental challenge lies in failure attribution: an observed failure may arise from the current model's behavior or from a systematic harness deficiency. The former leads to \emph{Model-Specific Accommodation} (MSA), whereas the latter calls for \emph{Harness-Level Repair} (HLR). Individual failures provide limited evidence for distinguishing these cases, while treating failures from a common harness issue as independent modification signals can also incur redundant modification and validation costs.
This distinction motivates a conceptual decomposition of harness adaptation. Consider the aggregate harness update $\Delta H$ produced over an evolution round, composed of a set of independent modification decisions issued by the coding agent. Each decision either accommodates a limitation of the current task model or repairs a systematic harness deficiency, contributing to $\Delta H_{\mathrm{MSA}}$ and $\Delta H_{\mathrm{HLR}}$, respectively. Letting $t\in[0,1]$ denote the proportion of these decisions that are MSA, the aggregate update decomposes as
\[
\Delta H =
t\,\Delta H_{\mathrm{MSA}} +
(1-t)\,\Delta H_{\mathrm{HLR}},
\qquad t\in[0,1],
\]
so that $t$ is at once the fraction of model-accommodating decisions and the weight of the model-accommodation component in the overall update. A larger $t$ may resolve observed failures efficiently but can overfit to the current model and training tasks. We hypothesize that failure patterns recurring across distinct task instances provide stronger evidence for systematic harness deficiencies than isolated failures, and therefore can shift adaptation away from unnecessary MSA. Our goal is not to develop a sophisticated evolution algorithm, but to use a simple and interpretable mechanism to validate our analysis that recurring failures across tasks provide a useful signal for harness-level evolution.

\subsection{Methodology}

\begin{algorithm}[t]
\caption{\textsc{Ecdysis} Training for Harness Self-Evolution}
\label{alg:harness-training}
\LinesNumbered
\KwIn{Fixed task model $\theta$, runtime environment $\mathcal{E}$, training task set $\mathcal{D}_{\mathrm{train}}$, initial harness $H_{\mathrm{base}}$, failure threshold $\lambda$, number of evolution rounds $R$, and number of refinement passes $K$}
\KwOut{Frozen final harness $H_F$}

$H_0\leftarrow H_{\mathrm{base}}$\;

\For{$i=1$ \KwTo $R$}{
  $\mathcal{T}_i\leftarrow
  \operatorname{Collect}(
  \theta,H_{i-1},\mathcal{E},\mathcal{D}_{\mathrm{train}})
  $
  \InlineComment{execution trajectories}\;

  $f_\lambda(\tau)\leftarrow
  \mathbb{I}\!\left[S(\tau)<\lambda\right]$\;

  $\mathcal{B}_i\leftarrow
  \operatorname{Aggregate}\!\left(
  \{\tau\in\mathcal{T}_i:f_\lambda(\tau)=1\}
  \right)$
  \InlineComment{structured failure evidence}\;

  \If{$\mathcal{B}_i=\varnothing$}{
    $H_i\leftarrow H_{i-1}$\;
    \textbf{continue}\;
  }

  $\mathcal{G}_i\leftarrow
  \operatorname{Group}(\mathcal{B}_i)$
  \InlineComment{failure patterns}\;

  $\mathcal{A}\leftarrow
  \{\operatorname{Analyst},\operatorname{Critic},\operatorname{Engineer}\}$\;
  $\mathcal{M}_i\leftarrow\varnothing$
  \InlineComment{shared role transcript}\;

  \For{$k=1$ \KwTo $K$}{
    \ForEach{$A\in\mathcal{A}$}{
      $m_i^{k,A}\leftarrow
      A(\mathcal{G}_i,\mathcal{M}_i)$\;
      $\mathcal{M}_i\leftarrow
      \mathcal{M}_i\mathbin{\circ}m_i^{k,A}$\;
    }
  }

  $q_i\leftarrow
  \operatorname{Moderator}
  (\mathcal{G}_i,\mathcal{M}_i)$
  \InlineComment{structured modification specification}\;

  $H_i^{\mathrm{c}}\leftarrow
  \operatorname{Edit}(H_{i-1},q_i)$
  \InlineComment{coding agent}\;

  \eIf{$J_{\mathrm{train}}(H_i^{\mathrm{c}})
  >
  J_{\mathrm{train}}(H_{i-1})$}{
    $H_i\leftarrow H_i^{\mathrm{c}}$\;
  }{
    $H_i\leftarrow H_{i-1}$\;
  }
}

$H_F\leftarrow H_R$\;
\KwRet{$H_F$}\;
\end{algorithm}

Given a task model $\theta$, a runtime environment $\mathcal{E}$, a training task set $\mathcal{D}_{\mathrm{train}}$, and an initial runtime harness $H_{\mathrm{base}}$, \textsc{Ecdysis} optimizes the harness using execution trajectories collected from the training tasks. Throughout training, the task model parameters and runtime environment remain fixed. We initialize the harness as $H_0=H_{\mathrm{base}}$ and denote by $H_i$ the validated harness retained after evolution round $i$.
For each evolution round $i\in\{1,2,\ldots,R\}$, \textsc{Ecdysis} proceeds in three stages. First, it executes the training tasks with the current harness $H_{i-1}$ and aggregates the resulting failure evidence. Second, it transforms the aggregated evidence into a structured modification specification, which is provided to a coding agent to modify $H_{i-1}$ and produce a candidate harness $H_i^{\mathrm{c}}$. Third, it validates $H_i^{\mathrm{c}}$ on the training set and retains it only if it improves the overall training score; otherwise, the previous harness is preserved. After $R$ rounds, training terminates and \textsc{Ecdysis} freezes the most recently validated harness, denoted by $H_F=H_R$. The complete procedure is summarized in Algorithm~\ref{alg:harness-training}.
Let $J_{\mathrm{train}}(H)$ denote the overall score of harness $H$ on the training set under the fixed evaluation framework. In evolution round $i$, \textsc{Ecdysis} accepts the candidate harness if and only if
$
J_{\mathrm{train}}\!\left(H_i^{\mathrm{c}}\right)
>
J_{\mathrm{train}}(H_{i-1}).
$
Accordingly,

$$
H_i=
\begin{cases}
H_i^{\mathrm{c}}, & \text{if } 
J_{\mathrm{train}}\!\left(H_i^{\mathrm{c}}\right)
>
J_{\mathrm{train}}(H_{i-1}),\\
H_{i-1}, & \text{otherwise}.
\end{cases}
$$

This criterion constrains only the overall training-set score and does not require every individual training task to improve. We next describe the two core components of \textsc{Ecdysis}: Batch-Level Failure Aggregation and Failure-Driven Collaborative Refinement (FDCR).

\subsection{Batch-Level Failure Aggregation}

At evolution round $i$, \textsc{Ecdysis} executes the training tasks using the current harness $H_{i-1}$ and collects the resulting execution trajectories, denoted by $\mathcal{T}_i$. A task specifies the target objective, whereas its execution trajectory records the concrete process leading to the observed outcome. The trajectory therefore provides execution-level evidence for failure analysis beyond the final task-level signal.

For each trajectory $\tau\in\mathcal{T}_i$, the fixed evaluation framework produces a task score $S(\tau)$. We define the binary failure signal as
$
f_\lambda(\tau)
=
\mathbb{I}\!\left[
S(\tau)<\lambda
\right]
$,
where $\lambda$ is a predefined threshold and $f_\lambda(\tau)=1$ indicates failure. Trajectories satisfying this condition are converted into structured records to form the failure evidence set $\mathcal{B}_i$. Each record retains the task identifier, failure decision, termination reason, tool-call history, and necessary execution context. This procedure uses the fixed evaluation framework as the training signal and does not modify or replace its scoring mechanism.
The failure records are organized into failure groups to facilitate the analysis of recurring patterns. \textsc{Ecdysis} prioritizes groups that cover at least two distinct tasks, since repeated failures from a single task alone do not establish that the underlying failure mechanism generalizes across tasks. Groups containing only one task identifier are nevertheless retained as auxiliary evidence for subsequent analysis. Importantly, failure groups serve as diagnostic evidence rather than mandatory repair targets. They help determine whether an observed pattern indicates a harness deficiency and guide decisions regarding the modification level, trigger conditions, modification scope, and safety constraints. Thus, a candidate harness need not address every observed failure group. Batch-level aggregation also reduces repeated coding-agent calls for candidate harness modification. Let $n_i=|\mathcal{B}_i|$ denote the number of failure records collected in round $i$. Processing failures independently requires \(N_{\mathrm{serial}}=\sum_{i=1}^{r}n_i\) coding-agent calls, whereas \textsc{Ecdysis} requires only \(N_{\mathrm{round}}=\sum_{i=1}^{r}\mathbb{I}(n_i>0)\) calls. Therefore, \(N_{\mathrm{round}}\leq N_{\mathrm{serial}}\), with strict inequality whenever any round contains multiple failure records.

\subsection{Failure-Driven Collaborative Refinement}

For each evolution round containing failure evidence, \textsc{Ecdysis} applies FDCR to transform the aggregated failure patterns into a structured harness modification specification. Inspired by collaborative criticism and iterative refinement in multi-agent systems \citep{yu-etal-2025-table, chen-etal-2025-magicore}, FDCR uses observed failure patterns as the driving evidence for role-based diagnosis and iterative refinement, rather than treating individual failures as independent modification requests.
FDCR separates failure analysis and modification planning from the actual harness implementation. Given the failure groups $\mathcal{G}_i$, the Analyst, Critic, and Engineer iteratively refine modification proposals through a shared transcript. The Analyst identifies potential harness deficiencies and proposes minimal, targeted updates. The Critic evaluates these proposals against the observed failure evidence and examines potential risks, including overly broad triggers, unintended blocking of legitimate behavior, violations of the runtime contract, and regressions on previously successful tasks. The Engineer tracks agreements and unresolved disagreements and identifies issues requiring clarification in subsequent refinement. The roles are executed sequentially, with each role receiving the accumulated transcript so that subsequent analysis can build on preceding discussion. The refinement proceeds for a fixed number of rounds.
After the role-based refinement, the Moderator reads the failure evidence and the complete transcript and produces a structured modification specification. The Moderator serves as a conservative arbitration stage that consolidates the refined proposals, resolves remaining disagreements, and prioritizes modifications that address recurring failure patterns while avoiding unnecessary changes. The resulting specification describes the identified failure patterns, proposed harness changes, and relevant implementation guidance, but does not directly modify the harness. Instead, the coding agent uses the specification together with the relevant execution evidence and the source code of the current harness to modify $H_{i-1}$ and produce the candidate harness $H_i^{\mathrm{c}}$. This separation keeps failure analysis structured and failure-driven while delegating implementation to the coding agent. By prioritizing failure patterns recurring across distinct task instances, FDCR biases the modification process toward systematic HLR and away from unnecessary MSA.

\section{Experiments}
\label{sec:experiments}
\subsection{Experimental Setup}
\label{sec:experimental-setup}
\noindent
\textbf{Models}. Our evaluation follows prior work \citep{xu2026adapting} and considers the compatibility between LLM reasoning capabilities and benchmark difficulty. We evaluate five task LLMs: Qwen3-8B, Qwen3-14B, Qwen3-32B \citep{yang2025qwen3technicalreport}, MiniMax-M2.7 (230B), and Llama-3.1-8B. For reproducibility, all LLMs are accessed through APIs. We use the same inference configuration for all task models in both training and held-out evaluation. For evaluation parameters, we refer to the baseline methods. The sampling temperature is set to 0.0. We use OpenCode\footnote{https://opencode.ai} with DeepSeek-V4-Pro as its underlying model as the coding agent. In \textsc{Ecdysis} (w/ FDCR), the Analyst, Critic, Engineer, and Moderator also use DeepSeek-V4-Pro \citep{deepseekai2026deepseekv4}.

\noindent
\textbf{Datasets}. To ensure comprehensive coverage of diverse task types, we use AgentBench \citep{liu2024agentbench} for relatively simple tasks and $\tau^2$-Bench for more complex and challenging tasks. Following prior work, we use the Airline and Retail subsets of $\tau^2$-Bench, which we refer to as $\tau^2$-Airline and $\tau^2$-Retail, respectively. Both subsets require agents to perform tool interactions that continuously modify the environment state \citep{barres2026taubench}.

\begin{figure}[t]
    \centering
    \includegraphics[width=\linewidth]{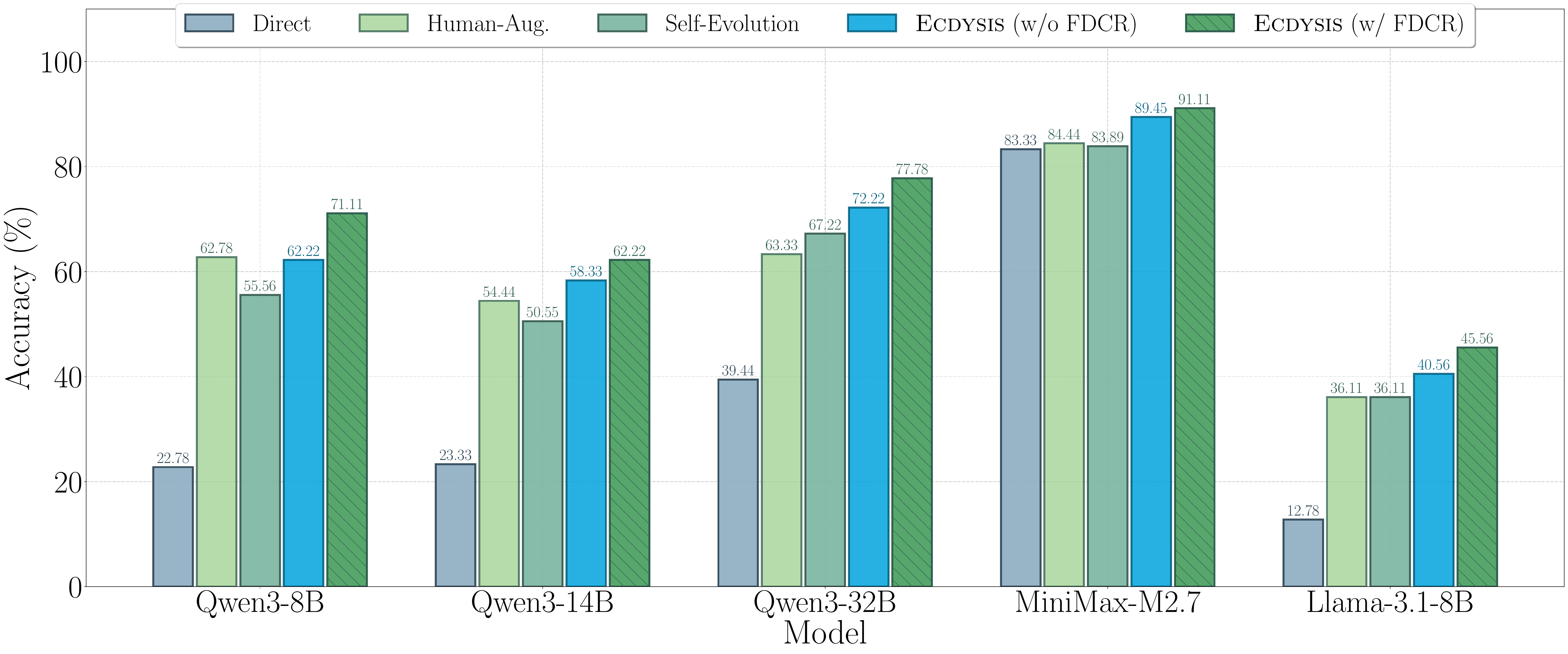}
    \caption{Overall accuracy of five harness self-evolution methods across task models and datasets.}
    \label{fig:overall-accuracy}
    \vspace{-15pt}
\end{figure}

\noindent
\textbf{Baselines and Ablation Study}.
For the runtime harness, we adopt \textit{Life-Harness}, a mature and well-structured harness framework \citep{xu2026adapting}. To enable fair and controlled comparisons, we construct five runtime harness configurations, covering a direct baseline, a human-optimized baseline, and three harness evolution strategies.

\begin{itemize}[left=0pt, itemsep=0pt]
\item \textbf{Direct}: retains the base agent loop, message handling, and tool interface required by $\tau^2$-Bench, while disabling the harness layers adopted from Life-Harness. This configuration serves as the minimal baseline without any additional harness mechanisms.

\item \textbf{Human-Augmented Harness (Human-Aug.)}: uses a fixed harness without harness evolution. This harness is obtained through human-involved optimization in prior work and serves as a strong baseline representing the performance of a manually optimized runtime harness. Moreover, the human-optimized harness is used as the common base harness for harness post-training. That is, all three evolution-based methods described below are initialized from the same Human-Aug. harness, thereby keeping the initial harness configuration fixed across methods.

\item \textbf{Self-Evolution (SE)}: performs instance-level serial updates. Each failure record independently invokes the coding agent, and the modifications produced for individual failure instances are applied sequentially and accumulated into a single round-level candidate. The accumulated candidate is then evaluated in the next complete training evaluation \citep{xu2026adapting}.

\item \textbf{\textsc{Ecdysis} (w/o FDCR)}: performs mixed training by aggregating failure evidence from multiple training tasks and evolution rounds. For each evolution round with non-empty failure evidence, it constructs a single round-level modification plan from the aggregated evidence and invokes the coding agent once to implement the proposed changes.

\item \textbf{\textsc{Ecdysis} (w/ FDCR)}: operates on the same mixed training input as \textsc{Ecdysis} (w/o FDCR). The Analyst, Critic, and Engineer agents jointly analyze the aggregated failure evidence through two rounds of FDCR. The Moderator then synthesizes the results and formulates the final modification plan, which is subsequently implemented by the coding agent.
\end{itemize}

Across all methods, only the runtime harness is updated during training, while the model parameters remain fixed. Thus, the comparison isolates the effect of different harness optimization strategies while controlling for the initial harness configuration and the underlying model. Due to the high complexity and cost of agent evaluation, we follow existing work \citep{qu2026she} in designing the experimental setup and scale. Additional experimental setup details are provided in Appendix \ref{app:exp_set}. We select Life-Harness's SE as our representative baseline, as it covers major runtime interventions while retaining the core principles of Self-Harness \citep{zhang2026self} and HarnessEvolve \citep{jiang2026harnessevolve}, while avoiding additional designs that could introduce confounding factors into our analysis. Moreover, our study isolates harness evolution under a frozen model, unlike approaches such as HarnessForge \citep{chen2026harnessforge} that jointly evolve the model and harness.

\section{Results}
\label{sec:results}

\noindent
\textbf{Overall Results}.
Across the fifteen combinations of five task models and three datasets, \textsc{Ecdysis} (w/ FDCR) achieves the highest average accuracy (see Table~\ref{tab:overall-results} and Figure~\ref{fig:overall-accuracy}). Its relative improvements over SE and Human-Aug. are 18.56\% and 15.51\%, respectively. It also improves Pass\textasciicircum3 by 28.56\% relative to SE, showing more consistent success across repeated trials. Beyond task accuracy, \textsc{Ecdysis} also improves both training and inference efficiency relative to SE. We report the training results and the inference results in Section~\ref{sec:results}, and analyze the sources of these gains in Section~\ref{sec:evolution-process-analysis}.

\begin{table}[t]
\centering
\caption{Overall task performance and inference efficiency across five LLMs and three datasets. Inference efficiency is measured by token consumption (M) and execution time (s).}
\label{tab:overall-results}
\scalebox{0.9}{%
\begin{tabular}{l|ccc|cc}
\toprule
\multirow{2}{*}{\textbf{Method}} &
    \multicolumn{3}{c|}{\textbf{Accuracy (\%)}} &
    \multicolumn{2}{c}{\textbf{Efficiency}} \\
\cmidrule(lr){2-4} \cmidrule(lr){5-6}
    & \textbf{AVG} & \textbf{Pass@3} & \textbf{Pass\textasciicircum3}
    & \textbf{Token Cost} & \textbf{Time} \\
\midrule
Direct
    & 36.33
    & 48.00
    & 23.67
    & $7.301\,\pm\,4.189$
    & $119.39\,\pm\,81.65$ \\
Human-Aug.
    & 60.22
    & 70.67
    & 49.33
    & $8.215\,\pm\,5.754$
    & $111.30\,\pm\,62.74$ \\
Self-Evolution
    & 58.67
    & 70.00
    & 46.67
    & $8.301\,\pm\,6.324$
    & $111.12\,\pm\,58.43$ \\
\textsc{Ecdysis} (w/o FDCR)
    & 64.56
    & 72.67
    & 56.00
    & $7.378\,\pm\,5.490$
    & $108.40\,\pm\,76.52$ \\
\textsc{Ecdysis} (w/ FDCR)
    & \textbf{69.56}
    & \textbf{77.67}
    & \textbf{60.00}
    & \textbf{$7.240\,\pm\,4.963$}
    & \textbf{$98.41\,\pm\,55.28$} \\
\bottomrule
\end{tabular}%
}
\vspace{-15pt}
\end{table}

\noindent
\textbf{Ablation Results}.
\label{sec:ablation-results}
All three task-performance metrics improve from SE to \textsc{Ecdysis} (w/o FDCR). The accuracy increases from 58.67\% to 64.56\%, an improvement of 5.89 percentage points. Meanwhile, Pass@3 and Pass\textasciicircum3 reach 72.67\% and 56.00\%, improving by 2.67 and 9.33 percentage points, respectively. Adding FDCR further increases the average task success rate to 69.56\%, Pass@3 to 77.67\%, and Pass\textasciicircum3 to 60.00\%, corresponding to additional gains of 5.00, 5.00, and 4.00 percentage points over \textsc{Ecdysis} (w/o FDCR). These results show that round-level aggregation of failure evidence across task instances provides performance gains on its own, with FDCR further improving the aggregate results.

\begin{wrapfigure}[15]{r}{0.5\textwidth}
\vspace{-10pt}
    \centering
    \includegraphics[width=\linewidth]{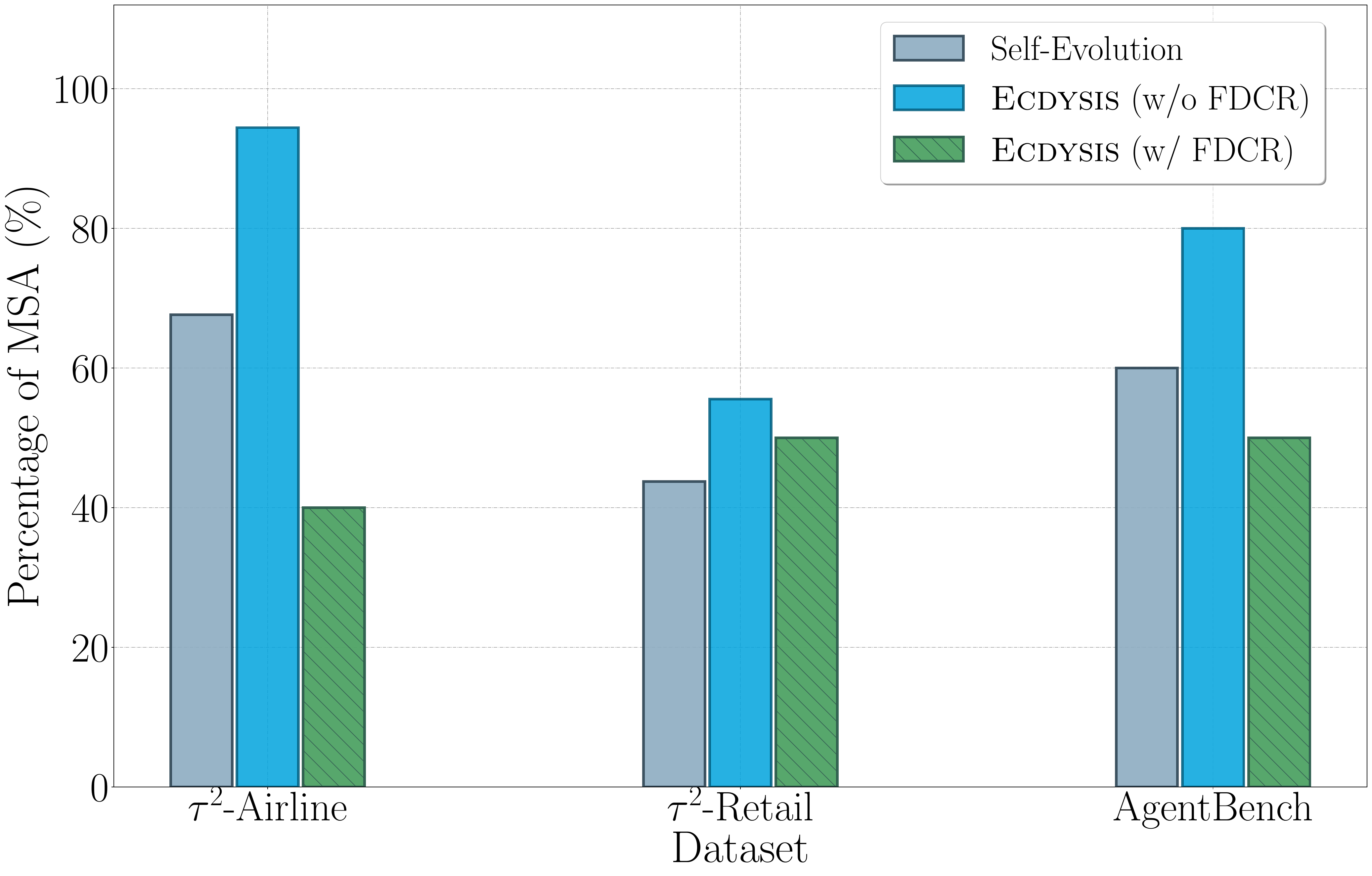}
    \caption{Model-accommodation ratios across datasets and harness evolution methods.}
    \label{fig:model-accommodation-ratios}
\end{wrapfigure}

\noindent
\textbf{Training Efficiency}.
The training-efficiency results show that round-level failure aggregation reduces repeated modification calls during harness evolution. \textsc{Ecdysis} (w/o FDCR) completes end-to-end training in 1,292.4 seconds on $\tau^2$-Retail and 2,510.8 seconds on $\tau^2$-Airline, corresponding to speedups of 1.42$\times$ and 3.23$\times$ over SE. \textsc{Ecdysis} (w/ FDCR) takes 1,405.9 and 4,403.0 seconds, corresponding to speedups of 1.30$\times$ and 1.84$\times$ (Table~\ref{tab:evolution-training-time}). API cost follows the same pattern. On $\tau^2$-Retail and $\tau^2$-Airline, SE costs \$8.484 and \$6.382, whereas \textsc{Ecdysis} (w/o FDCR) reduces these costs to \$2.485 and \$2.136, and \textsc{Ecdysis} (w/ FDCR) costs \$5.763 and \$2.609, respectively (see Table~\ref{tab:evolution-api-cost}). Overall, \textsc{Ecdysis} decouples efficient failure-driven evolution from costly collaborative refinement. Failure aggregation provides the primary efficiency gains, while FDCR serves as an accuracy-oriented refinement module that trades additional evolution-time cost for improved harness quality. Across the complete training process, SE, Ecdysis (w/o FDCR), and Ecdysis (w/ FDCR) made 881, 267, and 482 LLM calls, respectively. These results suggest that the gains from our schemes stem from our algorithmic design rather than increased LLM usage.

\noindent

\begin{table*}[t]
    \centering
    \caption{Inference cost comparison across five LLMs, three datasets, and five methods. Inference cost is measured by token consumption (M) and execution time (s).}
    \label{tab:heldout-inference-efficiency}
    \scalebox{0.65}{%
        \setlength{\tabcolsep}{4pt}%
        \begin{tabular}{l|l|cc|cc|cc}
        \toprule
        \multirow{2}{*}{\textbf{Model}} &
        \multirow{2}{*}{\textbf{Method}} &
        \multicolumn{2}{c|}{\textbf{$\tau^2$-Airline}} &
        \multicolumn{2}{c|}{\textbf{$\tau^2$-Retail}} &
        \multicolumn{2}{c}{\textbf{AgentBench}} \\
        \cmidrule(lr){3-4} \cmidrule(lr){5-6} \cmidrule(lr){7-8}
        & & \textbf{Tokens} & \textbf{Time} & \textbf{Tokens} & \textbf{Time} & \textbf{Tokens} & \textbf{Time} \\
        \midrule
\multirow{5}{*}{Qwen3-8B} & Direct & $9.544 \pm 0.352$ & $67.55 \pm 3.07$ & $8.470 \pm 0.416$ & $108.92 \pm 15.41$ & $5.615 \pm 0.073$ & $96.09 \pm 3.63$ \\
& Human-Aug. & $13.961 \pm 0.666$ & $108.58 \pm 15.42$ & $9.536 \pm 0.182$ & $113.88 \pm 2.25$ & $1.562 \pm 0.058$ & $46.67 \pm 0.73$ \\
& Self-Evolution & $12.321 \pm 1.205$ & $175.23 \pm 27.82$ & $9.483 \pm 0.495$ & $171.32 \pm 18.96$ & $1.418 \pm 0.094$ & $48.21 \pm 1.70$ \\
& \textsc{Ecdysis} (w/o FDCR) & $12.516 \pm 0.428$ & $102.43 \pm 1.59$ & $8.205 \pm 0.118$ & $180.13 \pm 20.16$ & $1.413 \pm 0.001$ & $47.24 \pm 0.70$ \\
\rowcolor{gray!15}\cellcolor{white}
& \textbf{\textsc{Ecdysis} (w/ FDCR)} & \textbf{$12.070 \pm 1.458$} & \textbf{$98.87 \pm 17.85$} & \textbf{$9.790 \pm 0.326$} & \textbf{$142.30 \pm 20.28$} & \textbf{$1.413 \pm 0.028$} & \textbf{$45.68 \pm 0.74$} \\
\midrule
\multirow{5}{*}{Qwen3-14B} & Direct & $13.685 \pm 0.858$ & $103.27 \pm 1.99$ & $19.239 \pm 2.460$ & $134.06 \pm 20.80$ & $4.853 \pm 0.089$ & $198.25 \pm 1.10$ \\
& Human-Aug. & $15.718 \pm 1.539$ & $101.83 \pm 19.15$ & $20.099 \pm 3.226$ & $158.45 \pm 42.92$ & $1.645 \pm 0.000$ & $52.01 \pm 0.06$ \\
& Self-Evolution & $24.308 \pm 0.420$ & $197.61 \pm 11.94$ & $14.392 \pm 3.268$ & $107.25 \pm 31.44$ & $1.618 \pm 0.001$ & $56.07 \pm 0.34$ \\
& \textsc{Ecdysis} (w/o FDCR) & $16.089 \pm 1.954$ & $157.02 \pm 31.85$ & $18.424 \pm 2.998$ & $128.51 \pm 49.22$ & $1.392 \pm 0.002$ & $51.77 \pm 1.12$ \\
\rowcolor{gray!15}\cellcolor{white}
& \textbf{\textsc{Ecdysis} (w/ FDCR)} & \textbf{$16.417 \pm 1.462$} & \textbf{$57.16 \pm 11.15$} & \textbf{$12.746 \pm 1.282$} & \textbf{$90.82 \pm 18.23$} & \textbf{$1.514 \pm 0.003$} & \textbf{$54.32 \pm 2.20$} \\
\midrule
\multirow{5}{*}{Qwen3-32B} & Direct & $6.469 \pm 0.349$ & $75.51 \pm 9.04$ & $7.182 \pm 0.066$ & $77.40 \pm 4.93$ & $4.330 \pm 0.183$ & $98.18 \pm 4.22$ \\
& Human-Aug. & $11.047 \pm 0.287$ & $105.05 \pm 8.83$ & $12.297 \pm 1.813$ & $88.55 \pm 11.70$ & $2.139 \pm 0.007$ & $61.74 \pm 0.26$ \\
& Self-Evolution & $13.206 \pm 0.483$ & $117.99 \pm 3.33$ & $9.023 \pm 0.353$ & $79.43 \pm 14.13$ & $1.789 \pm 0.026$ & $60.79 \pm 1.39$ \\
& \textsc{Ecdysis} (w/o FDCR) & $9.748 \pm 0.578$ & $91.34 \pm 9.55$ & $7.895 \pm 0.220$ & $73.45 \pm 1.11$ & $1.436 \pm 0.001$ & $54.40 \pm 1.12$ \\
\rowcolor{gray!15}\cellcolor{white}
& \textbf{\textsc{Ecdysis} (w/ FDCR)} & \textbf{$11.670 \pm 1.009$} & \textbf{$100.36 \pm 13.24$} & \textbf{$9.015 \pm 0.299$} & \textbf{$102.02 \pm 5.28$} & \textbf{$1.330 \pm 0.006$} & \textbf{$51.54 \pm 0.83$} \\
\midrule
\multirow{5}{*}{MiniMax-M2.7} & Direct & $5.252 \pm 0.171$ & $135.52 \pm 12.59$ & $5.173 \pm 0.048$ & $115.32 \pm 4.57$ & $2.937 \pm 0.444$ & $150.48 \pm 18.61$ \\
& Human-Aug. & $6.554 \pm 0.074$ & $154.62 \pm 3.26$ & $6.054 \pm 0.077$ & $270.38 \pm 40.65$ & $2.310 \pm 0.119$ & $162.26 \pm 10.01$ \\
& Self-Evolution & $8.161 \pm 0.327$ & $170.69 \pm 15.05$ & $5.728 \pm 0.052$ & $164.34 \pm 3.70$ & $2.066 \pm 0.065$ & $169.65 \pm 6.06$ \\
& \textsc{Ecdysis} (w/o FDCR) & $6.127 \pm 0.083$ & $139.25 \pm 3.80$ & $5.779 \pm 0.119$ & $335.37 \pm 43.90$ & $1.664 \pm 0.118$ & $114.02 \pm 4.39$ \\
\rowcolor{gray!15}\cellcolor{white}
& \textbf{\textsc{Ecdysis} (w/ FDCR)} & \textbf{$6.526 \pm 0.218$} & \textbf{$202.03 \pm 9.94$} & \textbf{$5.804 \pm 0.051$} & \textbf{$201.45 \pm 4.04$} & \textbf{$1.474 \pm 0.025$} & \textbf{$90.57 \pm 3.43$} \\
\midrule
\multirow{5}{*}{Llama-3.1-8B} & Direct & $5.818 \pm 0.360$ & $32.23 \pm 5.01$ & $6.181 \pm 0.266$ & $29.31 \pm 4.14$ & $4.771 \pm 0.170$ & $368.80 \pm 13.20$ \\
& Human-Aug. & $9.640 \pm 0.144$ & $51.68 \pm 3.22$ & $8.586 \pm 0.450$ & $32.32 \pm 5.33$ & $2.076 \pm 0.135$ & $161.42 \pm 7.92$ \\
& Self-Evolution & $10.319 \pm 0.299$ & $43.38 \pm 1.74$ & $8.724 \pm 0.276$ & $33.53 \pm 5.73$ & $1.952 \pm 0.197$ & $71.34 \pm 3.51$ \\
& \textsc{Ecdysis} (w/o FDCR) & $10.321 \pm 0.434$ & $51.97 \pm 4.52$ & $8.438 \pm 0.254$ & $54.72 \pm 1.40$ & $1.225 \pm 0.000$ & $44.43 \pm 0.48$ \\
\rowcolor{gray!15}\cellcolor{white}
& \textbf{\textsc{Ecdysis} (w/ FDCR)} & \textbf{$9.015 \pm 0.178$} & \textbf{$159.60 \pm 15.38$} & \textbf{$8.517 \pm 0.228$} & \textbf{$32.30 \pm 4.29$} & \textbf{$1.293 \pm 0.004$} & \textbf{$47.15 \pm 0.83$} \\
        \bottomrule
        \end{tabular}%
    }
\vspace{-15pt}
\end{table*}
\textbf{Inference Efficiency}.
We report token consumption and mean per-trajectory runtime for the complete held-out matrix in Table~\ref{tab:heldout-inference-efficiency}. Final evaluation tokens include only the input and output tokens consumed by the task model during the formal held-out trajectories. Runtime is measured per trajectory rather than as the wall-clock duration of the concurrent evaluation. Averaged over the 15 model and dataset combinations, final evaluation tokens are 8.301M for SE, 7.378M for \textsc{Ecdysis} (w/o FDCR), and 7.240M for \textsc{Ecdysis} (w/ FDCR), corresponding to relative reductions of 11.12\% and 12.78\%. This indicates that the efficiency acquired during evolution persists on held-out trajectories without any test-time modification. Runtime follows the same trend, with \textsc{Ecdysis} (w/ FDCR) reducing the mean from 111.12 to 98.41 seconds per trajectory. The largest gain is on Qwen3-14B over $\tau^2$-Airline, where the mean runtime drops from 197.61 to 57.16 seconds, corresponding to a 3.46$\times$ speedup. Because runtime also reflects model service latency and trajectory length, we report the full means and standard deviations in Table~\ref{tab:heldout-inference-efficiency} rather than only the aggregate.

\begin{table}[t]
    \centering
    \caption{Pooled input-cache utilization during harness evolution across three datasets.}
    \label{tab:pooled-evolution-cache-hit-rate}
    \scalebox{0.85}{%
        \setlength{\tabcolsep}{8pt}%
        \begin{tabular}{l|c|c|c}
        \toprule
        \textbf{Method} &
        \textbf{Cached Tokens} &
        \textbf{Total Input Tokens} &
        \textbf{Cache Hit Rate (\%)} \\
        \midrule
        Self-Evolution
            & 33,527,168
            & 37,326,451
            & 89.82 \\
        \textsc{Ecdysis} (w/o FDCR)
            & 15,881,728
            & 16,731,412
            & \textbf{94.92} \\
        \textsc{Ecdysis} (w/ FDCR)
            & 30,305,792
            & 31,935,263
            & 94.90 \\
        \bottomrule
        \end{tabular}%
    }
\end{table}
\begin{table}[t]
\centering
\caption{Training time (s) for the three harness self-evolution methods.}
\label{tab:evolution-training-time}
\scalebox{0.85}{%
    \setlength{\tabcolsep}{7pt}%
    \begin{tabular}{c|c|c|c|c}
    \toprule
    \textbf{Dataset} &
    \textbf{Self-Evolution} &
    \textbf{\textsc{Ecdysis} (w/o FDCR)} &
    \textbf{\textsc{Ecdysis} (w/ FDCR)} &
    \textbf{Speedup} \\
    \midrule
    $\tau^2$-Airline
        & 8,120.6
        & 2,510.8
        & 4,403.0
        & \textbf{3.23$\times$} / 1.84$\times$ \\
    $\tau^2$-Retail
        & 1,831.4
        & 1,292.4
        & 1,405.9
        & 1.42$\times$ / 1.30$\times$ \\
    AgentBench
        & 1,955.3
        & 1,136.6
        & 1,890.5
        & \textbf{1.72$\times$} / 1.03$\times$ \\
    \bottomrule
    \end{tabular}%
}
\end{table}

\section{Data Study}
\label{sec:task-structure-analysis}

\textbf{Motivation}. Recent work on on-policy distillation shows that training efficiency depends not only on the amount of training data, but also on the information contained in individual training instances. \citet{hou2026matters} find that a small set of carefully selected hard examples can nearly match training on a much larger dataset, with the gains attributed primarily to the longer reasoning trajectories induced by challenging problems. This raises a related question for harness evolution: \emph{which training instances provide the most informative signal for harness training?} Unlike model training, where longer reasoning trajectories can be particularly valuable, harness training requires sufficient task complexity and behavioral breadth to exercise diverse interaction structures and expose distinct failure mechanisms. We therefore view training tasks not merely as execution instances, but as diagnostic probes that reveal both the coverage of runtime behaviors and the deficiencies of the harness.

\subsection{Insights for Training Data Curation}
\label{sec:training-data-curation}
The patterns across tasks motivate a curation strategy based on failure information. Training tasks differ in diagnostic value. Some expose new execution paths or interaction structures. Some reveal harness deficiencies missed in earlier runs. Others repeat known failure mechanisms. Selecting training data only by quantity or surface diversity can be costly and may provide little new information for harness evolution. We therefore select training failures using interaction structures and failure mechanisms together. We compare the evidence across task instances. A failure enters the reduced training set when it adds a new path, interaction structure, or failure mechanism. A failure provides little new information when it repeats a signal already present in the selected set. This process also supports failure diagnosis. Failures from different tasks can be different forms of the same underlying harness deficiency. Aggregating these failures helps \textsc{Ecdysis} identify shared deficiencies and enable more general harness changes. To evaluate this strategy, we train \textsc{Ecdysis} with a reduced training set containing only five training tasks. We compare its performance with training on the full training set and with training on a randomly selected set of five tasks. All settings use the same evolution procedure and evaluation protocol. The reduced training set achieves performance comparable to the full training set while substantially reducing the training cost of harness evolution. These results suggest that the benefit comes from the information in the selected failures rather than from reducing the training set alone. Results are summarized in Table~\ref{tab:retail-data-curation}.

\begin{wrapfigure}[16]{r}{0.55\textwidth}
\vspace{-13pt}
    \centering
    \includegraphics[width=\linewidth]{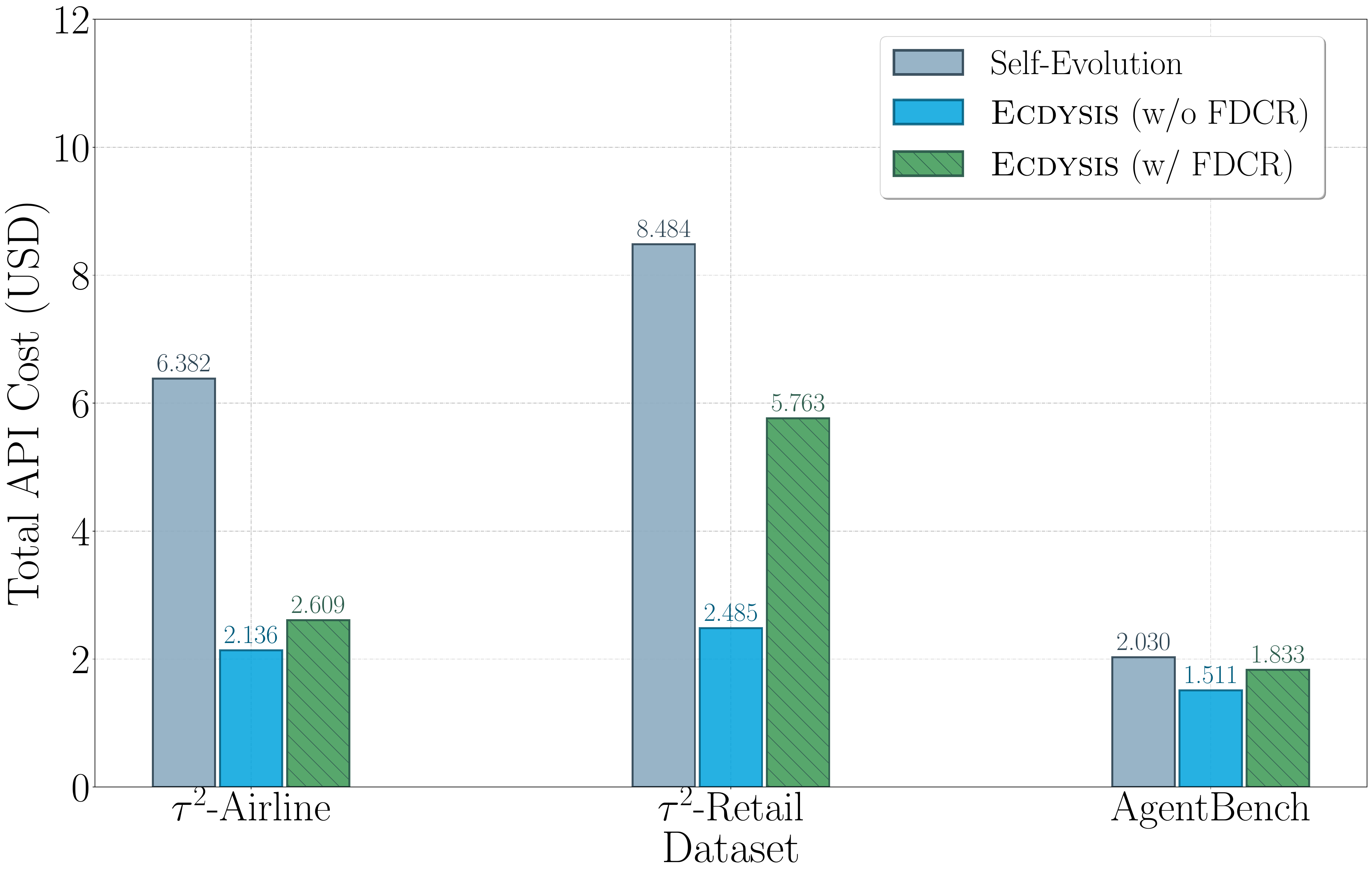}
    \caption{API cost of harness evolution across datasets and methods.}
    \label{fig:evolution-api-cost}
\end{wrapfigure}

\begin{table}[t]
\centering
\caption{
Data efficiency of harness evolution under different training configurations. Harnesses are evolved using Qwen3-8B and evaluated using Qwen3-32B.
}
\label{tab:retail-data-curation}
\scalebox{0.9}{%
    \setlength{\tabcolsep}{16pt}%
    \begin{tabular}{l|c|c|c|c}
    \toprule
    \textbf{Method} &
    \textbf{AVG (\%)} &
    \textbf{Pass@3} &
    \textbf{Pass\textasciicircum3} &
    \textbf{Training Cost} \\
    \midrule
    Self-Evolution
    & $65.00 \pm 5.00$ & 75.00 & 55.00 & \$8.484 \\
    \textsc{Ecdysis} (Full)
    & \textbf{75.00 $\pm$ 0.00} & 85.00
    & \textbf{60.00} & \$5.763 \\
    \textsc{Ecdysis} (Reduced)
    & $71.67 \pm 7.64$ & \textbf{90.00}
    & 45.00 & \textbf{\$0.291} \\
    \textsc{Ecdysis} (Random)
    & $65.00 \pm 8.66$ & 80.00
    & 45.00 & \$0.505 \\
    \bottomrule
    \end{tabular}%
}
\vspace{-15pt}
\end{table}

\subsection{Beyond Local Failure-Driven Evolution}
\label{sec:Beyond}
To better understand how failure-driven evolution modifies the runtime harness, we conducted a fine-grained manual analysis of the training-time data and quantified $t$ introduced in Section \ref{sec:preliminary-analysis}. Here, $t$ denotes the proportion of MSA. Local failure-driven evolution can overfit to task-specific outcomes by promoting a particular training-task answer into a general runtime constraint. In some cases, the evolution agent removes a legitimate action option to avoid a specific model error, replacing a conditional decision that should be made by the model with a global harness-level prohibition. Although such modifications may improve the triggering evaluation case, they shrink the valid action space of the harness and may impair its generalization across models.
We manually audited substantive harness modifications recorded in the saved round artifacts, including candidates later rolled back, while excluding temporary, unpersisted attempts. An independent modification decision is classified as either MSA or HLR. We compute $t$ by pooling independent modification decisions across the three datasets. We observe $t = 60.0\%$ for SE, compared with $t = 45.5\%$ for \textsc{Ecdysis}. Cross-instance aggregation alone does not automatically reduce MSA: \textsc{Ecdysis} without FDCR has a pooled ratio of 75.61\%, whereas FDCR reduces it to 45.45\% (see Figure~\ref{fig:model-accommodation-ratios}). This indicates that the reduction is primarily associated with the combination of cross-instance failure analysis and FDCR. These results suggest that \textsc{Ecdysis} mitigates rather than eliminates MSA, directing persistent modifications toward evidence-supported runtime deficiencies while preserving a broader valid action space and reducing dependence on the model used during evolution. Detailed audit criteria, dataset-level results, and qualitative cases are provided in Appendix~\ref{app:model-accommodation-analysis}.

\section{Conclusion}

In this paper, we present \textsc{Ecdysis}, a framework for failure-driven evolution of runtime harnesses for LLM agents. Our central insight is that execution failures may reflect model limitations rather than harness deficiencies, making individual failures unreliable modification signals. \textsc{Ecdysis} therefore separates failure-driven evolution into two complementary problems: cross-task failure recurrence for efficiently selecting evidence worth acting on, and FDCR for attributing recurring failures before translating them into harness modifications. Recurrence improves evolution efficiency and data efficiency, but may over-amplify recurring model-specific failures. FDCR incurs additional computation while reducing such accommodation and improving cross-LLM generalization. Across multiple LLMs and benchmarks, \textsc{Ecdysis} improves harness performance and reduces inference-time token consumption. Overall, effective harness evolution requires both efficient evidence selection and reliable failure attribution to produce modifications that generalize beyond the models and tasks used during evolution.

\section*{Ethical Considerations}
In this paper, AI assistants are used to polish the writing. We also use AI agents to assist with programming. All AI-assisted outputs are reviewed and verified by the authors. All human experts involved in the manual annotation experiments are authors of this paper. No external human participants are involved in the annotation process.

\bibliographystyle{unsrtnat}
\bibliography{neurips_2025}

@inproceedings{
shao2026your,
title={Your Agent May Misevolve: Emergent Risks in Self-evolving {LLM} Agents},
author={Shuai Shao and Qihan Ren and Dongrui Liu and Chen Qian and Boyi Wei and Dadi Guo and Yang JingYi and Xinhao Song and Linfeng Zhang and Weinan Zhang and Jing Shao},
booktitle={The Fourteenth International Conference on Learning Representations},
year={2026},
url={https://openreview.net/forum?id=Fd1jgQQW28}
}

@article{jiang2026self,
  title={Self-Improving Agents in the Era of Experience: A Survey of Self-to Meta-Evolution},
  author={Jiang, Che and Zhong, Jincheng and Fu, Yu and Tian, Kai and Yang, Junlin and Zhao, Kaikai and Wang, Yuchong and Luo, Tianwei and Wang, Weizhi and Zuo, Yuxin and others},
  year={2026}
}

@article{chen2026failed,
  title={From Failed Trajectories to Reliable LLM Agents: Diagnosing and Repairing Harness Flaws},
  author={Chen, Mengzhuo and Wang, Junjie and Liu, Zhe and Wang, Yawen and Zheng, Haiming and Wang, Qing},
  journal={arXiv preprint arXiv:2606.06324},
  year={2026}
}

@article{xu2026adapting,
  title={Adapting the interface, not the model: Runtime harness adaptation for deterministic llm agents},
  author={Xu, Tianshi and Wen, Huifeng and Li, Meng},
  journal={arXiv preprint arXiv:2605.22166},
  year={2026}
}

@inproceedings{
barres2026taubench,
title={$\tau^2$-Bench: Evaluating Conversational Agents in a Dual-Control Environment},
author={Victor Barres and Honghua Dong and Soham Ray and Xujie Si and Karthik R Narasimhan},
booktitle={Forty-third International Conference on Machine Learning},
year={2026},
url={https://openreview.net/forum?id=OC2z7iSQKa}
}

@misc{yang2025qwen3technicalreport,
      title={Qwen3 Technical Report}, 
      author={An Yang and Anfeng Li and Baosong Yang and Beichen Zhang and Binyuan Hui and Bo Zheng and Bowen Yu and Chang Gao and Chengen Huang and Chenxu Lv and Chujie Zheng and Dayiheng Liu and Fan Zhou and Fei Huang and Feng Hu and Hao Ge and Haoran Wei and Huan Lin and Jialong Tang and Jian Yang and Jianhong Tu and Jianwei Zhang and Jianxin Yang and Jiaxi Yang and Jing Zhou and Jingren Zhou and Junyang Lin and Kai Dang and Keqin Bao and Kexin Yang and Le Yu and Lianghao Deng and Mei Li and Mingfeng Xue and Mingze Li and Pei Zhang and Peng Wang and Qin Zhu and Rui Men and Ruize Gao and Shixuan Liu and Shuang Luo and Tianhao Li and Tianyi Tang and Wenbiao Yin and Xingzhang Ren and Xinyu Wang and Xinyu Zhang and Xuancheng Ren and Yang Fan and Yang Su and Yichang Zhang and Yinger Zhang and Yu Wan and Yuqiong Liu and Zekun Wang and Zeyu Cui and Zhenru Zhang and Zhipeng Zhou and Zihan Qiu},
      year={2025},
      eprint={2505.09388},
      archivePrefix={arXiv},
      primaryClass={cs.CL},
      url={https://arxiv.org/abs/2505.09388}, 
}

@misc{deepseekai2026deepseekv4,
      title={DeepSeek-V4: Towards Highly Efficient Million-Token Context Intelligence}, 
      author={Anyi Xu and Bangcai Lin and Bing Xue and Bingxuan Wang and Bingzheng Xu and Bochao Wu and Bowei Zhang and Chaofan Lin and Chen Dong and Chenchen Ling and Chengda Lu and Chenggang Zhao and Chengqi Deng and Chengyu Hou and Chenhao Xu and Chenze Shao and Chong Ruan and Conner Sun and Damai Dai and Daya Guo and Dejian Yang and Deli Chen and Donghao Li and Dongjie Ji and Erhang Li and Fang Wei and Fangyun Lin and Fangzhou Yuan and Feiyu Xia and Fucong Dai and Guangbo Hao and Guanting Chen and Guoai Cao and Guolai Meng and Guowei Li and Han Yu and Han Zhang and Hanwei Xu and Hao Li and Haofen Liang and Haoling Zhang and Haoming Luo and Haoran Wei and Haotian Yuan and Haowei Zhang and Haowen Luo and Haoyu Chen and Haozhe Ji and Hengqing Zhang and Honghui Ding and Hongxuan Tang and Huanqi Cao and Huazuo Gao and Hui Qu and Hui Zeng and J Yang and JQ Zhu and Jia Luo and Jia Song and Jia Yu and Jialiang Huang and Jialu Cai and Jian Liang and Jiangting Zhou and Jiasheng Ye and Jiashi Li and Jiaxin Xu and Jiewen Hu and Jieyu Yang and Jin Chen and Jin Yan and Jingchang Chen and Jingli Zhou and Jingting Xiang and Jingyang Yuan and Jingyuan Cheng and Jingzi Zhou and Jinhua Zhu and Jiping Yu and Joseph Sun and Jun Ran and Junguang Jiang and Junjie Qiu and Junlong Li and Junmin Zheng and Junxiao Song and Kai Dong and Kaige Gao and Kang Guan and Kexing Zhou and Kezhao Huang and Kuai Yu and Lean Wang and Lecong Zhang and Lei Wang and Leyi Xia and Li Zhang and Liang Zhao and Lihua Guo and Lingxiao Luo and Linwang Ma and Linyan Zhu and Litong Wang and Liyu Cai and Liyue Zhang and Longhao Chen and MS Di and MY Xu and Max Mei and Miaojun Wang and Mingchuan Zhang and Minghua Zhang and Minghui Tang and Mingming Li and Mingxu Zhou and Minmin Han and Ning Wang and Panpan Huang and Panpan Wang and Peixin Cong and Peiyi Wang and Peng Zhang and Qiancheng Wang and Qihao Zhu and Qingyang Li and Qinyu Chen and Qiushi Du and Qiwei Jiang and Rui Tian and Ruifan Xu and Ruijie Lu and Ruiling Xu and Ruiqi Ge and Ruisong Zhang and Ruizhe Pan and Runji Wang and Runqian Chen and Runqiu Yin and Runxin Xu and Ruomeng Shen and Ruoyu Zhang and Ruyi Chen and SH Liu and Shanghao Lu and Shangmian Sun and Shangyan Zhou and Shanhuang Chen and Shaofei Cai and Shaoheng Nie and Shaoqing Wu and Shaoyuan Chen and Shengding Hu and Shengyu Liu and Shiqiang Hu and Shirong Ma and Shiyu Wang and Shuiping Yu and Shunfeng Zhou and Shuting Pan and Shuying Yu and Songyang Zhou and Tao Ni and Tao Yun and Tian Jin and Tian Pei and Tian Ye and Tianle Lin and Tianran Ji and Tianyi Cui and Tianyuan Yue and Tingting Yu and Tun Wang and W Zhang and WL Xiao and Wangding Zeng and Wei An and Weilin Zhao and Wen Liu and Wenfeng Liang and Wenjie Pang and Wenjing Luo and Wenjing Yao and Wenjun Gao and Wenkai Yang and Wenlve Huang and Wenqing Hou and Wentao Zhang and Wenting Ma and Xi Gao and Xiang He and Xiangwen Wang and Xianzu Wang and Xiao Bi and Xiaodong Liu and Xiaohan Wang and Xiaokang Chen and Xiaokang Zhang and Xiaotao Nie and Xiaowen Sun and Xiaoxiang Wang and Xin Cheng and Xin Liu and Xin Xie and Xingchao Liu and Xingchen Liu and Xingkai Yu and Xingyou Li and Xinyu Yang and Xinyu Zhang and Xu Chen and Xuanyu Wang and Xuecheng Su and Xueyin Chen and Xuheng Lin and Xuwei Fu and YC Yan and YQ Wang and YW Ma and Yanfeng Luo and Yang Zhang and Yanhong Xu and Yanru Ma and Yanwen Huang and Yao Li and Yao Li and Yao Xu and Yao Zhao and Yaofeng Sun and Yaohui Wang and Yi Qian and Yi Shao and Yi Yu and Yichao Zhang and Yifan Ding and Yifan Shi and Yijia Wu and Yiliang Xiong and Yiling Ma and Ying He and Ying Tang and Ying Zhou and Yingjia Luo and Yinmin Zhong and Yishi Piao and Yisong Wang and Yixiang Zhang and Yixiao Chen and Yixuan Tan and Yixuan Wei and Yiyang Ma and Yiyuan Liu and Yonglun Yang and Yongqiang Guo and Yongtong Wu and Yu Wu and YuKun Li and Yuan Cheng and Yuan Ou and Yuanfan Xu and Yuanhao Li and Yuduan Wang and Yuehan Yang and Yuer Xu and Yuhan Wu and Yuhao Meng and Yuheng Zou and Yukun Zha and Yunfan Xiong and Yupeng Chen and Yuping Lin and Yuqian Cao and Yuqian Wang and Yushun Zhang and Yuting Yan and Yutong Lin and Yuxian Gu and Yuxiang Luo and Yuxiang You and Yuxuan Liu and Yuxuan Zhou and Yuyang Zhou and Yuzhen Huang and ZF Wu and Zehao Wang and Zehua Zhao and Zehui Ren and Zekai Zhang and Zhangli Sha and Zhe Fu and Zhe Ju and Zhean Xu and Zhenda Xie and Zhengyan Zhang and Zheren Gao and Zhewen Hao and Zhibin Gou and Zhicheng Ma and Zhigang Yan and Zhihong Shao and Zhixian Huang and Zhixuan Chen and Zhiyu Wu and Zhizhou Ren and Zhongyu Wu and Zhuoshu Li and Zhuping Zhang and Zian Xu and Zihao Wang and Zihua Qu and Zihui Gu and Zijia Zhu and Zilin Li and Zipeng Zhang and Ziwei Xie and Ziyi Gao and Ziyi Wan and Zizheng Pan and Zongqing Yao},
      year={2026},
      eprint={2606.19348},
      archivePrefix={arXiv},
      primaryClass={cs.CL},
      url={https://arxiv.org/abs/2606.19348}, 
}

@inproceedings{
wang2026huxleygodel,
title={Huxley-G\"odel Machine: Human-Level Coding Agent Development by an Approximation of the Optimal Self-Improving Machine},
author={Wenyi Wang and Piotr Pi{\k{e}}kos and Li Nanbo and Firas Laakom and Yimeng Chen and Mateusz Ostaszewski and Mingchen Zhuge and J{\"u}rgen Schmidhuber},
booktitle={The Fourteenth International Conference on Learning Representations},
year={2026},
url={https://openreview.net/forum?id=T0EiEuhOOL}
}

@misc{zhang2026harnesscompass,
      title={HarnessCompass: Guiding Automatic Harness Evolution toward Generalizable and Effective Agent Harnesses}, 
      author={Luan Zhang and Ruochen Zhou and Dandan Song and Zhengyu Chen and Yuhang Tian and Jun Yang and Huipeng Ma and Chenhao Li and Guangyuan Feng and Xudong Li and Yizhou Jin and Yan Xu},
      year={2026},
      eprint={2608.01918},
      archivePrefix={arXiv},
      primaryClass={cs.LG},
      url={https://arxiv.org/abs/2608.01918}, 
}

@misc{qu2026she,
      title={SHE: Trajectory-driven Safety Harness Evolution for LLM Agents}, 
      author={Wanying Qu and Qinghua Mao and Yu Li and Jiyao Liu and Xin Zhang and Dadi Guo and Yanxu Zhu and Qingyu Liu and Leitao Yuan and Xi Lin and Shanfeng Zhu and Yanwei Fu and Jing Shao and Xia Hu and Dongrui Liu},
      year={2026},
      eprint={2608.09885},
      archivePrefix={arXiv},
      primaryClass={cs.AI},
      url={https://arxiv.org/abs/2608.09885}, 
}

@inproceedings{yu-etal-2025-table,
    title = "Table-Critic: A Multi-Agent Framework for Collaborative Criticism and Refinement in Table Reasoning",
    author = "Yu, Peiying  and
      Chen, Guoxin  and
      Wang, Jingjing",
    editor = "Che, Wanxiang  and
      Nabende, Joyce  and
      Shutova, Ekaterina  and
      Pilehvar, Mohammad Taher",
    booktitle = "Proceedings of the 63rd Annual Meeting of the Association for Computational Linguistics (Volume 1: Long Papers)",
    month = jul,
    year = "2025",
    address = "Vienna, Austria",
    publisher = "Association for Computational Linguistics",
    url = "https://aclanthology.org/2025.acl-long.853/",
    doi = "10.18653/v1/2025.acl-long.853",
    pages = "17432--17451",
    ISBN = "979-8-89176-251-0"
}

@inproceedings{
wang2026rethinking,
title={Rethinking the Evaluation of Harness Evolution for Agents},
author={Yike Wang and Huaisheng Zhu and Zhengyu Hu and Yige Yuan and Zhengyu Chen and Shakti Senthil and Hannaneh Hajishirzi and Yulia Tsvetkov and Pradeep Dasigi and Teng Xiao},
booktitle={COLM 2026 The 2nd Workshop on Lifelong Agents: Learning, Aligning, and Evolving},
year={2026},
url={https://openreview.net/forum?id=WVAeeSlVim}
}

@inproceedings{chen-etal-2025-locagent,
    title = "{L}oc{A}gent: Graph-Guided {LLM} Agents for Code Localization",
    author = "Chen, Zhaoling  and
      Tang, Robert  and
      Deng, Gangda  and
      Wu, Fang  and
      Wu, Jialong  and
      Jiang, Zhiwei  and
      Prasanna, Viktor  and
      Cohan, Arman  and
      Wang, Xingyao",
    editor = "Che, Wanxiang  and
      Nabende, Joyce  and
      Shutova, Ekaterina  and
      Pilehvar, Mohammad Taher",
    booktitle = "Proceedings of the 63rd Annual Meeting of the Association for Computational Linguistics (Volume 1: Long Papers)",
    month = jul,
    year = "2025",
    address = "Vienna, Austria",
    publisher = "Association for Computational Linguistics",
    url = "https://aclanthology.org/2025.acl-long.426/",
    doi = "10.18653/v1/2025.acl-long.426",
    pages = "8697--8727",
    ISBN = "979-8-89176-251-0"
}

@article{shao2026harness,
  title={Harness-R1: Learning to Edit Executable Runtime Harnesses from Agent Failure Trajectories},
  author={Shao, Shuai and Zhang, Kangning and Li, Qingyao and Wang, Shijian and Wang, Hao and Jiao, Wenxiang and Lu, Yuan and Guo, Yi and Liu, Weiwen and Zhang, Weinan},
  journal={arXiv preprint arXiv:2608.02276},
  year={2026}
}

@inproceedings{
yao2023react,
title={ReAct: Synergizing Reasoning and Acting in Language Models},
author={Shunyu Yao and Jeffrey Zhao and Dian Yu and Nan Du and Izhak Shafran and Karthik R Narasimhan and Yuan Cao},
booktitle={The Eleventh International Conference on Learning Representations },
year={2023},
url={https://openreview.net/forum?id=WE_vluYUL-X}
}

@inproceedings{wan-etal-2026-compass,
    title = "{COMPASS}: Enhancing Agent Long-Horizon Reasoning with Evolving Context",
    author = "Wan, Guangya  and
      Ling, Mingyang  and
      Ren, Xiaoqi  and
      Han, Rujun  and
      Li, Sheng  and
      Zhang, Zizhao",
    editor = "Liakata, Maria  and
      Moreira, Viviane P.  and
      Zhang, Jiajun  and
      Jurgens, David",
    booktitle = "Proceedings of the 64th Annual Meeting of the {A}ssociation for {C}omputational {L}inguistics (Volume 1: Long Papers)",
    month = jul,
    year = "2026",
    address = "San Diego, California, United States",
    publisher = "Association for Computational Linguistics",
    url = "https://aclanthology.org/2026.acl-long.152/",
    doi = "10.18653/v1/2026.acl-long.152",
    pages = "3360--3380",
    ISBN = "979-8-89176-390-6"
}

@inproceedings{shinn2023ref,
 author = {Shinn, Noah and Cassano, Federico and Gopinath, Ashwin and Narasimhan, Karthik and Yao, Shunyu},
 booktitle = {Advances in Neural Information Processing Systems},
 doi = {10.52202/075280-0377},
 editor = {A. Oh and T. Naumann and A. Globerson and K. Saenko and M. Hardt and S. Levine},
 pages = {8634--8652},
 publisher = {Curran Associates, Inc.},
 title = {Reflexion: language agents with verbal reinforcement learning},
 url = {https://proceedings.neurips.cc/paper_files/paper/2023/file/1b44b878bb782e6954cd888628510e90-Paper-Conference.pdf},
 volume = {36},
 year = {2023}
}

@inproceedings{
khattab2024dspy,
title={{DSP}y: Compiling Declarative Language Model Calls into State-of-the-Art Pipelines},
author={Omar Khattab and Arnav Singhvi and Paridhi Maheshwari and Zhiyuan Zhang and Keshav Santhanam and Sri Vardhamanan A and Saiful Haq and Ashutosh Sharma and Thomas T. Joshi and Hanna Moazam and Heather Miller and Matei Zaharia and Christopher Potts},
booktitle={The Twelfth International Conference on Learning Representations},
year={2024},
url={https://openreview.net/forum?id=sY5N0zY5Od}
}

@inproceedings{yang2024large,
 author = {Yang, Chengrun and Wang, Xuezhi and Lu, Yifeng and Liu, Hanxiao and Le, Quoc V and Zhou, Denny and Chen, Xinyun},
 booktitle = {International Conference on Learning Representations},
 editor = {B. Kim and Y. Yue and S. Chaudhuri and K. Fragkiadaki and M. Khan and Y. Sun},
 pages = {12028--12068},
 title = {Large Language Models as Optimizers},
 url = {https://proceedings.iclr.cc/paper_files/paper/2024/file/3339f19c5fcee3ad74502947a32be9e6-Paper-Conference.pdf},
 volume = {2024},
 year = {2024}
}

@inproceedings{hu2025adas,
 author = {Hu, Shengran and Lu, Cong and Clune, Jeff},
 booktitle = {International Conference on Learning Representations},
 editor = {Y. Yue and A. Garg and N. Peng and F. Sha and R. Yu},
 pages = {21344--21377},
 title = {Automated Design of Agentic Systems},
 url = {https://proceedings.iclr.cc/paper_files/paper/2025/file/36b7acf6f6010652b3f2a433774a66fe-Paper-Conference.pdf},
 volume = {2025},
 year = {2025}
}

@inproceedings{zhang2025aflow,
 author = {Zhang, Jiayi and Xiang, Jinyu and Yu, Zhaoyang and Teng, Fengwei and Chen, XiongHui and Chen, Jiaqi and Zhuge, Mingchen and Cheng, Xin and Hong, Sirui and Wang, Jinlin and Zheng, Bingnan and Liu, Bang and Luo, Yuyu and Wu, Chenglin},
 booktitle = {International Conference on Learning Representations},
 editor = {Y. Yue and A. Garg and N. Peng and F. Sha and R. Yu},
 pages = {34040--34077},
 title = {AFlow: Automating Agentic Workflow Generation},
 url = {https://proceedings.iclr.cc/paper_files/paper/2025/file/5492ecbce4439401798dcd2c90be94cd-Paper-Conference.pdf},
 volume = {2025},
 year = {2025}
}

@inproceedings{
zhang2026darwin,
title={Darwin G\"odel Machine: Open-Ended Evolution of Self-Improving Agents},
author={Jenny Zhang and Shengran Hu and Cong Lu and Robert Tjarko Lange and Jeff Clune},
booktitle={The Fourteenth International Conference on Learning Representations},
year={2026},
url={https://openreview.net/forum?id=pUpzQZTvGY}
}

@misc{zhang2026self,
      title={Self-Harness: Harnesses That Improve Themselves}, 
      author={Hangfan Zhang and Shao Zhang and Kangcong Li and Chen Zhang and Yang Chen and Yiqun Zhang and Lei Bai and Shuyue Hu},
      year={2026},
      eprint={2606.09498},
      archivePrefix={arXiv},
      primaryClass={cs.CL},
      url={https://arxiv.org/abs/2606.09498}, 
}

@misc{lee2026recursive,
      title={Recursive Harness Self-Improvement}, 
      author={Hyunin Lee and Jinglue Xu and Jeffrey Seely and Donghyun Lee and Matei Zaharia and Yujin Tang},
      year={2026},
      eprint={2607.15524},
      archivePrefix={arXiv},
      primaryClass={cs.LG},
      url={https://arxiv.org/abs/2607.15524}, 
}

@article{chen2026harnessforge,
  title={Harnessforge: Joint harness and policy evolution for adaptive agent systems},
  author={Chen, Mingju and Lv, Can and Zhang, Guibin and Chang, Heng and Zhou, Shiji},
  journal={arXiv preprint arXiv:2606.01779},
  year={2026}
}

@misc{wang2026handbook,
      title={Harness Handbook: Making Evolving Agent Harnesses Readable,Navigable, and Editable}, 
      author={Ruhan Wang and Yucheng Shi and Zongxia Li and Zhongzhi Li and Yue Yu and Junyao Yang and Kishan Panaganti and Haitao Mi and Dongruo Zhou and Leoweiliang},
      year={2026},
      eprint={2607.13285},
      archivePrefix={arXiv},
      primaryClass={cs.AI},
      url={https://arxiv.org/abs/2607.13285}, 
}

@misc{huang2026memo,
      title={MemoHarness: Agent Harnesses That Learn from Experience}, 
      author={Yue Huang and Wenjie Wang and Han Bao and Yuchen Ma and Xiaonan Luo and Yi Nian and Haomin Zhuang and Zheyuan Liu and Yue Zhao and Xiangliang Zhang},
      year={2026},
      eprint={2607.14159},
      archivePrefix={arXiv},
      primaryClass={cs.AI},
      url={https://arxiv.org/abs/2607.14159}, 
}

@misc{hou2026matters,
      title={What Matters in On-Policy Distillation? A Perspective on Data Efficiency and Data Selection}, 
      author={Zhinan Hou and Jiaqi Zhang and Xunliang Cai and Keyou You},
      year={2026},
      eprint={2609.05198},
      archivePrefix={arXiv},
      primaryClass={cs.AI},
      url={https://arxiv.org/abs/2609.05198}, 
}

@article{jiang2026harnessevolve,
  title={HarnessEvolve: Learning from Reference Trajectories for Reliable Agent Self-Evolution},
  author={Jiang, Wen and Chu, Mingmin and Tian, Yimeng and Zhang, Qianxin and Yang, Haofei and Yang, Rui and Liu, Yang and Lv, Tao and Li, Fangming},
  journal={arXiv preprint arXiv:2609.00829},
  year={2026}
}

@article{xu2026verify,
  title={Verify Smarter, Evolve Further: Efficient Harness Evolution through Behavior-Aware Verification},
  author={Xu, Jinghan and Zhang, Yikai and Chen, Aili and Li, Weiyuan and Liang, Jiaqing and Yang, Deqing},
  journal={arXiv preprint arXiv:2608.27311},
  year={2026}
}

@inproceedings{chen-etal-2025-magicore,
    title = "{MA}g{IC}o{R}e: Multi-Agent, Iterative, Coarse-to-Fine Refinement for Reasoning",
    author = "Chen, Justin  and
      Prasad, Archiki  and
      Saha, Swarnadeep  and
      Stengel-Eskin, Elias  and
      Bansal, Mohit",
    editor = "Christodoulopoulos, Christos  and
      Chakraborty, Tanmoy  and
      Rose, Carolyn  and
      Peng, Violet",
    booktitle = "Proceedings of the 2025 Conference on Empirical Methods in Natural Language Processing",
    month = nov,
    year = "2025",
    address = "Suzhou, China",
    publisher = "Association for Computational Linguistics",
    url = "https://aclanthology.org/2025.emnlp-main.1660/",
    doi = "10.18653/v1/2025.emnlp-main.1660",
    pages = "32663--32686",
    ISBN = "979-8-89176-332-6"
}

@inproceedings{zhang-etal-2025-enhancing-recommendation,
    title = "Enhancing Recommendation Explanations through User-Centric Refinement",
    author = "Zhang, Jingsen  and
      Tian, Zihang  and
      Feng, Xueyang  and
      Chen, Xu  and
      Chen, Chong",
    editor = "Christodoulopoulos, Christos  and
      Chakraborty, Tanmoy  and
      Rose, Carolyn  and
      Peng, Violet",
    booktitle = "Findings of the Association for Computational Linguistics: EMNLP 2025",
    month = nov,
    year = "2025",
    address = "Suzhou, China",
    publisher = "Association for Computational Linguistics",
    url = "https://aclanthology.org/2025.findings-emnlp.434/",
    doi = "10.18653/v1/2025.findings-emnlp.434",
    pages = "8177--8191",
    ISBN = "979-8-89176-335-7"
}

@inproceedings{
liu2024agentbench,
title={AgentBench: Evaluating {LLM}s as Agents},
author={Xiao Liu and Hao Yu and Hanchen Zhang and Yifan Xu and Xuanyu Lei and Hanyu Lai and Yu Gu and Hangliang Ding and Kaiwen Men and Kejuan Yang and Shudan Zhang and Xiang Deng and Aohan Zeng and Zhengxiao Du and Chenhui Zhang and Sheng Shen and Tianjun Zhang and Yu Su and Huan Sun and Minlie Huang and Yuxiao Dong and Jie Tang},
booktitle={The Twelfth International Conference on Learning Representations},
year={2024},
url={https://openreview.net/forum?id=zAdUB0aCTQ}
}


\newpage
\appendix
\definecolor{promptgreen}{rgb}{0,0.5,0.2}
\definecolor{diffremoved}{RGB}{160,38,38}
\definecolor{diffadded}{RGB}{0,110,55}

\tcbset{appendixbox/.style={breakable,enhanced,colback=green!3,colframe=promptgreen,colbacktitle=promptgreen,coltitle=white,fonttitle=\bfseries\footnotesize,fontupper=\footnotesize,fontlower=\footnotesize,left=.03in,right=.03in,bottom=.03in,top=.03in,before upper={\setlength{\baselineskip}{14pt}}}}
\newtcolorbox{promptbox}[1]{appendixbox,title={#1}}
\newtcolorbox{evidencebox}[1]{appendixbox,title={#1}}

\section{Manual Analysis}
\label{app:model-accommodation-analysis}
We conducted a fine-grained manual audit of the training-time harness modifications to characterize how observed failures are ultimately translated into modification decisions. The audit covers substantive modifications recorded in the saved round artifacts, including candidates that were later rolled back, while excluding temporary attempts that were never persisted. The unit of analysis is an \emph{independent modification decision}, defined by its target problem and direct behavioral change. Multiple edits serving the same purpose are merged into a single decision, even when they span multiple functions, registration points, or supporting skills. Conversely, distinct issues within the same function are counted separately. For each decision, we examine the failure trajectory, the information available to the model and its actual actions, the task requirements and environment rules, the coding agent's modification record, and the behavior before and after modification. We then inspect which triggering conditions and available actions are changed. Thus, classification cannot be inferred from the failure score or implementation form alone: a prompt modification is not necessarily MSA, and a code-level rule is not necessarily an HLR. Each independent modification decision is assigned to one of two categories:

\begin{itemize}[left=0pt, itemsep=0pt]
\item \textbf{MSA}: the modification primarily compensates for limitations of the current task model in understanding, planning, action selection, execution, or stopping decisions through prompting, recovery steps, additional constraints, action gating, or related mechanisms, rather than correcting a confirmed harness deficiency. Such support can be effective and reusable across tasks. Its classification does not imply inevitable overfitting.
\item \textbf{HLR}: environmental rules or execution evidence indicate an error or omission in parsing, state maintenance, interface semantics, execution conditions, or feedback presentation, and the modification directly repairs the runtime mechanism rather than deciding which task-level action the model should take.
\end{itemize}
We define the MSA ratio as the number of MSA decisions divided by all independent modification decisions. We further validate the MSA/HLR classification with five human experts, who independently evaluate each instance. The final labels are determined by majority voting, providing additional evidence for the reliability of the reported MSA ratios. The Overall result is computed by pooling decisions across $\tau^2$-Retail, $\tau^2$-Airline, and AgentBench rather than taking an arithmetic mean of dataset-level ratios. The ratio varies across datasets, indicating that modification tendencies depend on task structure and failure mechanisms rather than on a fixed harness layer or method label. For SE, the pooled ratio is 60.00\%, compared with 45.45\% for \textsc{Ecdysis} (w/ FDCR). This corresponds to a reduction of 14.55 percentage points. On $\tau^2$-Airline and AgentBench, the reductions are 27.65 and 10.00 percentage points, respectively, whereas $\tau^2$-Retail increases slightly from 43.75\% to 50.00\%. Nevertheless, \textsc{Ecdysis} with FDCR is lower than its w/o FDCR counterpart on all three datasets.

The ablation isolates the contributions of cross-instance aggregation and FDCR. \textsc{Ecdysis} without FDCR has higher MSA ratios than SE on all three datasets, showing that aggregation alone does not automatically reduce MSA. Adding FDCR reduces the ratio on every dataset: from 55.56\% to 50.00\% on $\tau^2$-Retail, from 94.44\% to 40.00\% on $\tau^2$-Airline, and from 80.00\% to 50.00\% on AgentBench. When all independent modification decisions are pooled, the ratio decreases from 75.61\% to 45.45\%, a reduction of 30.16 percentage points. This pattern is consistent with the roles of the two components: cross-instance aggregation organizes related failure evidence, while FDCR further scrutinizes the failure evidence, modification basis, triggering conditions, and potential collateral effects before a harness change is committed. The results therefore indicate that the reduction in MSA is primarily associated with their combination rather than aggregation alone.

The qualitative cases illustrate why the distinction requires evidence-based analysis. In a $\tau^2$-Retail trajectory, the model completed a return but selected a gift card when the user did not specify a refund destination, whereas the evaluation target was the original payment method. The environment policy nevertheless permits a gift-card refund when explicitly requested. The coding agent globally restricted refunds to the original payment method, preventing the observed error but eliminating a policy-permitted option. This is classified as \emph{MSA}. In another Retail modification, the prompt stated that a modified order could still be canceled, while the cancellation tool rejected such requests. The modification aligned the prompt with the actual execution semantics and is therefore an HLR.

A second HLR occurred in $\tau^2$-Airline. The task model used a write operation to query reservation information and submitted a baggage count identical to the current value. Although the baggage state remained unchanged, the operation created an unnecessary payment-history record that could affect subsequent refund calculations. The coding agent therefore added a safeguard that detects and rejects such no-op writes while directing the model to use a read operation instead. This modification prevents unintended state changes rather than prescribing a task-level action and is therefore classified as HLR.

These cases show that the same failure-driven process can yield different modification categories. The Retail refund case converts a local model error into a global restriction, whereas the other Retail and Airline cases correct erroneous runtime feedback or prevent harmful execution side effects. Consequently, neither failure outcomes nor prompt/code implementation forms can substitute for examining the underlying evidence, environment contract, and behavioral scope. The analysis thus provides the classification basis for the model-accommodation ratio and explains why \textsc{Ecdysis} combines cross-task failure analysis with FDCR: to direct persistent modifications toward evidence-supported runtime deficiencies while avoiding unnecessary conversion of local model errors into global constraints.

\section{Supporting Evidence for the Qualitative Cases}
\label{app:qualitative-case-evidence}

We provide the supporting policy and code excerpts for the three qualitative cases discussed in the preceding section: the Retail refund restriction, the Retail cancellation-feedback correction, and the Airline no-op baggage-write safeguard. The excerpts show the relevant rules and actual modifications underlying their classification as MSA or HLR.

\begin{evidencebox}{Refund restriction: Self-Evolution on $\tau^2$-Retail}
\textbf{Existing policy.}

\begin{quote}
The user needs to provide a payment method to receive the refund.

The refund must either go to the original payment method, or an existing gift card.
\end{quote}

\textbf{Harness diff.}

\begin{lstlisting}[basicstyle=\ttfamily\scriptsize,columns=fullflexible,keepspaces=true,showstringspaces=false,breaklines=true,breakatwhitespace=false,aboveskip=5pt,belowskip=5pt,moredelim={[l][\color{diffremoved}]{-}},moredelim={[l][\color{diffadded}]{+}}]
-        if not isinstance(pm, GiftCard) and payment_method_id != original_pm_id:
+        if payment_method_id != original_pm_id:
             raise ValueError(
                 ...
             )
\end{lstlisting}
\end{evidencebox}

The removed condition allowed an existing gift card. The added condition rejects every refund destination other than the original payment method. This is the policy-permitted option excluded by the refund restriction described in the preceding section.

\begin{evidencebox}{Cancellation feedback: Self-Evolution on $\tau^2$-Retail}
\textbf{Existing cancellation check.}

\begin{lstlisting}[basicstyle=\ttfamily\scriptsize,columns=fullflexible,keepspaces=true,showstringspaces=false,breaklines=true,breakatwhitespace=false,aboveskip=5pt,belowskip=5pt]
        if st == "pending (item modified)":
            raise ValueError(
                f"Order {order_id} has status 'pending (item modified)'. "
                "After items have been modified once, this order can no longer be cancelled "
                "with cancel_pending_order."
            )
\end{lstlisting}

\textbf{Harness diff.}

\begin{lstlisting}[basicstyle=\ttfamily\scriptsize,columns=fullflexible,keepspaces=true,showstringspaces=false,breaklines=true,breakatwhitespace=false,aboveskip=5pt,belowskip=5pt,moredelim={[l][\color{diffremoved}]{-}},moredelim={[l][\color{diffadded}]{+}}]
                 f"Note: Order {oid} items have already been modified once "
                 "(status: 'pending (item modified)'). Items cannot be modified "
-                "a second time. Address or payment method changes and full-order "
-                "cancellation are still available."
+                "a second time. Address or payment method changes are still "
+                "available, but this order CANNOT be cancelled with "
+                "cancel_pending_order."
\end{lstlisting}
\end{evidencebox}

The existing check already rejected cancellation after an item modification. The diff changes the feedback to match that check rather than adding a new cancellation restriction. This is a separate modification decision from the Retail refund restriction.

\begin{evidencebox}{No-op baggage-write protection: Self-Evolution on $\tau^2$-Airline}
\textbf{Failure described in the coding record.}

The coding record reports that the model used \texttt{update\_reservation\_baggages} to inspect a reservation and submitted unchanged baggage counts. It identifies an unintended payment-history entry from this no-op write as a source of errors in later cancellation refunds.

\textbf{Harness diff.}

\begin{lstlisting}[basicstyle=\ttfamily\scriptsize,columns=fullflexible,keepspaces=true,showstringspaces=false,breaklines=true,breakatwhitespace=false,aboveskip=5pt,belowskip=5pt,moredelim={[l][\color{diffremoved}]{-}},moredelim={[l][\color{diffadded}]{+}}]
+class NoOpBaggageUpdateRule:
     ...
+    tool_name = "update_reservation_baggages"
     ...
+        r = db.reservations.get(reservation_id)
+        if r is None:
+            return
+        if (total_baggages == r.total_baggages
+                and nonfree_baggages == r.nonfree_baggages):
+            raise ValueError(
                 ...
+                "This update would create a no-op payment entry that alters the "
+                "reservation's payment history. "
                 ...
             )
\end{lstlisting}
\end{evidencebox}

The added check blocks the write only when both requested baggage counts equal their current values. This is the no-op write safeguard described in the preceding section, addressing the payment-history side effect identified in the coding record.

\section{Additional Experimental Setup}
\label{app:exp_set}
\textbf{Models}. We evaluate five task LLMs: Qwen3-8B, Qwen3-14B, Qwen3-32B \citep{yang2025qwen3technicalreport}, MiniMax-M2.7 (230B)\footnote{https://huggingface.co/MiniMaxAI/MiniMax-M2.7}, and Llama-3.1-8B\footnote{https://huggingface.co/meta-llama/Llama-3.1-8B}.

\noindent
\textbf{Evaluation Protocol}. 
The three methods share the same training-time evaluation and candidate acceptance protocol, differing only in how they organize failure evidence and construct candidate modifications. We use Qwen3-8B as the task model and allow at most three candidate-generation rounds. Starting from the same initial harness, each method generates candidates from the resulting evidence, which are evaluated only in the subsequent complete evaluation and never on the trajectories used for their generation. A candidate is retained only if it strictly improves over the previously retained harness; otherwise, it is discarded. Accepted candidates are used for subsequent generation. After the final candidate is generated, a final evaluation over all training tasks is performed solely to determine whether it is accepted, with no further candidate generation. All evolution decisions are based exclusively on the training split. After evolution, the resulting harnesses are frozen and evaluated on held-out data under two independent protocols: first, Qwen3-8B is evaluated on all held-out tasks in each subset with three trials per task; second, we evaluate the full matrix of five task models, three datasets, and five methods, directly reusing the harnesses evolved with Qwen3-8B for the three evolution-based methods without further evolution. 

\noindent
\textbf{Evaluation Metrics}.
We evaluate \textsc{Ecdysis} from three perspectives: inference performance, training efficiency, and cost.
\begin{itemize}[left=0pt, itemsep=0pt]

\item \textbf{Inference Performance}.
On the held-out split, we report \textit{average task accuracy}, \textit{Pass@3}, and \textit{Pass\textasciicircum3}. Pass@3 denotes the proportion of tasks solved in at least one of the three trials, whereas Pass\textasciicircum3 denotes the proportion solved in all three trials.

\item \textbf{Training Efficiency}.
\textit{Training Time} measures the end-to-end wall-clock time from the initial training evaluation until the final harness is frozen. For held-out evaluation, we report \textit{Final Evaluation Tokens}, defined as the total input and output tokens consumed by the task model.

\item \textbf{Cost}.
We report \textit{Input Cache Hit Rate} and \textit{API Cost}. Input Cache Hit Rate is the ratio of cached input tokens to total input tokens. API Cost is reported separately for training task evaluation and harness modification.

\end{itemize}

\section{Detailed Experimental Results}
\label{app:detailed-results}

This appendix provides detailed results underlying the aggregate comparisons in the main text. We report per-dataset held-out task performance (Tables~\ref{tab:heldout-domain-accuracy} and~\ref{tab:agentbench-accuracy}), dataset-level evolution-efficiency statistics (Tables~\ref{tab:evolution-cache-utilization} and~\ref{tab:evolution-api-cost}), and MSA ratios (Table~\ref{tab:model-accommodation-analysis}). The main text focuses on pooled comparisons, whereas the tables below show dataset-level variation and detailed cost breakdowns.

\subsection{Task Performance}

Averaged over the five task models and the three datasets ($\tau^2$-Airline, $\tau^2$-Retail, and AgentBench), the average accuracy increases from 58.67\% under SE to 69.56\% with \textsc{Ecdysis} (w/ FDCR), a relative gain of 18.56\%. Results on $\tau^2$-Airline and $\tau^2$-Retail are detailed in Table~\ref{tab:heldout-domain-accuracy}, and those on AgentBench are reported in Table~\ref{tab:agentbench-accuracy}. The improvement is especially pronounced for Qwen3-8B on the $\tau^2$-Airline subset, where accuracy rises from 35.00\% to 60.00\%, Pass@3 from 50.00\% to 80.00\%, and Pass\textasciicircum3 from 20.00\% to 40.00\%. Notably, this improvement extends to models not used during evolution. On the same subset, Qwen3-32B improves from 51.67\% under SE to 68.33\% with \textsc{Ecdysis}. These results demonstrate that a harness evolved with Qwen3-8B transfers to other task models without further evolution.

\subsection{Held-Out Task Performance}

Table~\ref{tab:heldout-domain-accuracy} reports the detailed held-out results on the two $\tau^2$-Bench subsets, while Table~\ref{tab:agentbench-accuracy} reports the corresponding results on AgentBench. These tables provide the detailed values underlying the aggregate performance reported in Table~\ref{tab:overall-results}.

\begin{table*}[t]
\centering
\caption{Held-out task performance across five LLMs and two datasets.}
\label{tab:heldout-domain-accuracy}
\scalebox{0.84}{
    \setlength{\tabcolsep}{12pt}
    \begin{tabular}{l|l|l|ccc}
    \toprule
    \multirow{2}{*}{\textbf{Model}} & \multirow{2}{*}{\textbf{Dataset}} & \multirow{2}{*}{\textbf{Method}} & \multicolumn{3}{c}{\textbf{Accuracy (\%)}} \\
    \cmidrule(lr){4-6}
    & & & \textbf{AVG} & \textbf{Pass@3} & \textbf{Pass\textasciicircum3} \\
    \midrule
    \multirow{10}{*}{Qwen3-8B} & \multirow{5}{*}{$\tau^2$-Airline} & Direct & 16.67 $\pm$ 7.64 & 35.00 & 0.00 \\
    & & Human-Aug. & 55.00 $\pm$ 5.00 & 65.00 & 40.00 \\
    & & Self-Evolution & 35.00 $\pm$ 10.00 & 50.00 & 20.00 \\
    & & \textsc{Ecdysis} (w/o FDCR) & 48.33 $\pm$ 5.77 & 55.00 & \textbf{45.00} \\
    & & \textsc{Ecdysis} (w/ FDCR) & \textbf{60.00 $\pm$ 5.00} & \textbf{80.00} & 40.00 \\
    \cmidrule(lr){2-6}
    & \multirow{5}{*}{$\tau^2$-Retail} & Direct & 41.67 $\pm$ 5.77 & 70.00 & 15.00 \\
    & & Human-Aug. & 48.33 $\pm$ 2.89 & 80.00 & 20.00 \\
    & & Self-Evolution & 41.67 $\pm$ 15.28 & 75.00 & 15.00 \\
    & & \textsc{Ecdysis} (w/o FDCR) & 53.33 $\pm$ 2.89 & 75.00 & 25.00 \\
    & & \textsc{Ecdysis} (w/ FDCR) & \textbf{63.33 $\pm$ 7.64} & \textbf{85.00} & \textbf{45.00} \\
    \midrule
    \multirow{10}{*}{Qwen3-14B} & \multirow{5}{*}{$\tau^2$-Airline} & Direct & 16.67 $\pm$ 7.64 & 30.00 & 0.00 \\
    & & Human-Aug. & 38.33 $\pm$ 10.41 & 55.00 & 25.00 \\
    & & Self-Evolution & 23.33 $\pm$ 7.64 & 40.00 & 10.00 \\
    & & \textsc{Ecdysis} (w/o FDCR) & 41.67 $\pm$ 7.64 & 60.00 & \textbf{25.00} \\
    & & \textsc{Ecdysis} (w/ FDCR) & \textbf{43.33 $\pm$ 7.64} & \textbf{65.00} & \textbf{25.00} \\
    \cmidrule(lr){2-6}
    & \multirow{5}{*}{$\tau^2$-Retail} & Direct & 25.00 $\pm$ 5.00 & 50.00 & 5.00 \\
    & & Human-Aug. & 45.00 $\pm$ 13.23 & 65.00 & 20.00 \\
    & & Self-Evolution & 43.33 $\pm$ 12.58 & 65.00 & 20.00 \\
    & & \textsc{Ecdysis} (w/o FDCR) & 48.33 $\pm$ 5.77 & \textbf{70.00} & \textbf{35.00} \\
    & & \textsc{Ecdysis} (w/ FDCR) & \textbf{53.33 $\pm$ 2.89} & \textbf{70.00} & 30.00 \\
    \midrule
    \multirow{10}{*}{Qwen3-32B} & \multirow{5}{*}{$\tau^2$-Airline} & Direct & 20.00 $\pm$ 5.00 & 35.00 & 10.00 \\
    & & Human-Aug. & 56.67 $\pm$ 14.43 & 70.00 & 40.00 \\
    & & Self-Evolution & 51.67 $\pm$ 11.55 & \textbf{80.00} & 20.00 \\
    & & \textsc{Ecdysis} (w/o FDCR) & 61.67 $\pm$ 5.77 & \textbf{80.00} & 40.00 \\
    & & \textsc{Ecdysis} (w/ FDCR) & \textbf{68.33 $\pm$ 2.89} & \textbf{80.00} & \textbf{50.00} \\
    \cmidrule(lr){2-6}
    & \multirow{5}{*}{$\tau^2$-Retail} & Direct & 58.33 $\pm$ 2.89 & 75.00 & 35.00 \\
    & & Human-Aug. & 58.33 $\pm$ 7.64 & 80.00 & 45.00 \\
    & & Self-Evolution & 65.00 $\pm$ 5.00 & 75.00 & 55.00 \\
    & & \textsc{Ecdysis} (w/o FDCR) & 70.00 $\pm$ 8.66 & \textbf{85.00} & 45.00 \\
    & & \textsc{Ecdysis} (w/ FDCR) & \textbf{75.00 $\pm$ 0.00} & \textbf{85.00} & \textbf{60.00} \\
    \midrule
    \multirow{10}{*}{MiniMax-M2.7} & \multirow{5}{*}{$\tau^2$-Airline} & Direct & 80.00 $\pm$ 5.00 & \textbf{95.00} & 60.00 \\
    & & Human-Aug. & 81.67 $\pm$ 5.77 & 90.00 & 70.00 \\
    & & Self-Evolution & 73.33 $\pm$ 2.89 & 90.00 & 45.00 \\
    & & \textsc{Ecdysis} (w/o FDCR) & 81.67 $\pm$ 2.89 & 85.00 & \textbf{80.00} \\
    & & \textsc{Ecdysis} (w/ FDCR) & \textbf{86.67 $\pm$ 7.64} & \textbf{95.00} & 70.00 \\
    \cmidrule(lr){2-6}
    & \multirow{5}{*}{$\tau^2$-Retail} & Direct & 90.00 $\pm$ 5.00 & \textbf{100.00} & 80.00 \\
    & & Human-Aug. & 93.33 $\pm$ 5.77 & \textbf{100.00} & 85.00 \\
    & & Self-Evolution & 95.00 $\pm$ 5.00 & \textbf{100.00} & 90.00 \\
    & & \textsc{Ecdysis} (w/o FDCR) & \textbf{100.00 $\pm$ 0.00} & \textbf{100.00} & \textbf{100.00} \\
    & & \textsc{Ecdysis} (w/ FDCR) & 96.67 $\pm$ 2.89 & \textbf{100.00} & 90.00 \\
    \midrule
    \multirow{10}{*}{Llama-3.1-8B} & \multirow{5}{*}{$\tau^2$-Airline} & Direct & 25.00 $\pm$ 8.66 & 35.00 & 15.00 \\
    & & Human-Aug. & 31.67 $\pm$ 7.64 & 40.00 & 20.00 \\
    & & Self-Evolution & 28.33 $\pm$ 10.41 & \textbf{45.00} & 10.00 \\
    & & \textsc{Ecdysis} (w/o FDCR) & 31.67 $\pm$ 5.77 & 40.00 & 20.00 \\
    & & \textsc{Ecdysis} (w/ FDCR) & \textbf{35.00 $\pm$ 5.00} & 40.00 & \textbf{30.00} \\
    \cmidrule(lr){2-6}
    & \multirow{5}{*}{$\tau^2$-Retail} & Direct & 8.33 $\pm$ 2.89 & 10.00 & 5.00 \\
    & & Human-Aug. & 8.33 $\pm$ 2.89 & \textbf{15.00} & 5.00 \\
    & & Self-Evolution & 10.00 $\pm$ 0.00 & \textbf{15.00} & 5.00 \\
    & & \textsc{Ecdysis} (w/o FDCR) & 10.00 $\pm$ 5.00 & \textbf{15.00} & 5.00 \\
    & & \textsc{Ecdysis} (w/ FDCR) & \textbf{11.67 $\pm$ 2.89} & \textbf{15.00} & \textbf{10.00} \\
    \bottomrule
    \end{tabular}
}
\end{table*}

\begin{table*}[t]
\centering
\caption{Accuracy results across five models on AgentBench.}
\label{tab:agentbench-accuracy}
\scalebox{0.84}{%
    \setlength{\tabcolsep}{12pt}%
    \begin{tabular}{l|l|ccc}
    \toprule
    \multirow{2}{*}{\textbf{Model}}
    & \multirow{2}{*}{\textbf{Method}}
    & \multicolumn{3}{c}{\textbf{Accuracy (\%)}} \\
    \cmidrule(lr){3-5}
    & & \textbf{AVG} & \textbf{Pass@3} & \textbf{Pass\textasciicircum3} \\
    \midrule
    
    \multirow{5}{*}{Qwen3-8B}
    & Direct & 10.00 & 15.00 & 5.00 \\
    & Human-Aug. & 85.00 & \textbf{90.00} & 80.00 \\
    & Self-Evolution & \textbf{90.00} & \textbf{90.00} & \textbf{90.00} \\
    & \textsc{Ecdysis} (w/o FDCR) & 85.00 & 85.00 & 85.00 \\
    & \textsc{Ecdysis} (w/ FDCR)
    & \textbf{90.00} & \textbf{90.00} & \textbf{90.00} \\
    \midrule
    
    \multirow{5}{*}{Qwen3-14B}
    & Direct & 28.33 & 30.00 & 25.00 \\
    & Human-Aug. & 80.00 & 80.00 & 80.00 \\
    & Self-Evolution & 85.00 & 85.00 & 85.00 \\
    & \textsc{Ecdysis} (w/o FDCR) & 85.00 & 85.00 & 85.00 \\
    & \textsc{Ecdysis} (w/ FDCR)
    & \textbf{90.00} & \textbf{90.00} & \textbf{90.00} \\
    \midrule
    
    \multirow{5}{*}{Qwen3-32B}
    & Direct & 40.00 & 45.00 & 30.00 \\
    & Human-Aug. & 75.00 & 75.00 & 75.00 \\
    & Self-Evolution & 85.00 & 85.00 & 85.00 \\
    & \textsc{Ecdysis} (w/o FDCR) & 85.00 & 85.00 & 85.00 \\
    & \textsc{Ecdysis} (w/ FDCR)
    & \textbf{90.00} & \textbf{90.00} & \textbf{90.00} \\
    \midrule
    
    \multirow{5}{*}{MiniMax-M2.7}
    & Direct & 80.00 & \textbf{90.00} & 65.00 \\
    & Human-Aug. & 78.33 & 85.00 & 70.00 \\
    & Self-Evolution & 83.33 & 85.00 & 80.00 \\
    & \textsc{Ecdysis} (w/o FDCR) & 86.67 & \textbf{90.00} & 85.00 \\
    & \textsc{Ecdysis} (w/ FDCR)
    & \textbf{90.00} & \textbf{90.00} & \textbf{90.00} \\
    \midrule
    
    \multirow{5}{*}{Llama-3.1-8B}
    & Direct & 5.00 & 5.00 & 5.00 \\
    & Human-Aug. & 68.33 & 70.00 & 65.00 \\
    & Self-Evolution & 70.00 & 70.00 & 70.00 \\
    & \textsc{Ecdysis} (w/o FDCR) & 80.00 & 80.00 & 80.00 \\
    & \textsc{Ecdysis} (w/ FDCR)
    & \textbf{90.00} & \textbf{90.00} & \textbf{90.00} \\
    \bottomrule
    \end{tabular}%
    }
\end{table*}

\subsection{Evolution Efficiency}

Table~\ref{tab:evolution-cache-utilization} provides dataset-level cache-utilization statistics, while Table~\ref{tab:evolution-api-cost} reports the API-cost breakdown between training evaluation and evolution. These tables complement the pooled cache-utilization table (Table~\ref{tab:pooled-evolution-cache-hit-rate}) and API-cost figure (Fig.~\ref{fig:evolution-api-cost}) in the main text.

\begin{table}[t]
    \centering
    \caption{Input-cache utilization during harness evolution across three datasets.}
    \label{tab:evolution-cache-utilization}
    \scalebox{0.85}{%
        \setlength{\tabcolsep}{8pt}%
        \begin{tabular}{c|l|c|c|c}
        \toprule
        \textbf{Dataset} &
        \textbf{Method} &
        \textbf{Cached Tokens} &
        \textbf{Total Input Tokens} &
        \textbf{Cache Hit Rate (\%)} \\
        \midrule

        \multirow{3}{*}{\textbf{$\tau^2$-Airline}}
        & Self-Evolution & 12,368,896 & 13,824,385 & 89.47 \\
        & \textsc{Ecdysis} (w/o FDCR) & 4,384,768 & 4,601,409 & \textbf{95.29} \\
        & \textsc{Ecdysis} (w/ FDCR) & 5,103,104 & 5,470,490 & 93.28 \\
        
        \midrule

        \multirow{3}{*}{\textbf{$\tau^2$-Retail}}
        & Self-Evolution & 16,375,808 & 18,327,058 & 89.35 \\
        & \textsc{Ecdysis} (w/o FDCR) & 7,570,944 & 7,885,487 & \textbf{96.01} \\
        & \textsc{Ecdysis} (w/ FDCR) & 19,547,136 & 20,485,995 & 95.42 \\
        
        \midrule

        \multirow{3}{*}{\textbf{AgentBench}}
        & Self-Evolution & 4,782,464 & 5,175,008 & 92.41 \\
        & \textsc{Ecdysis} (w/o FDCR) & 3,926,016 & 4,244,516 & 92.50 \\
        & \textsc{Ecdysis} (w/ FDCR) & 5,655,552 & 5,978,778 & \textbf{94.59} \\

        \bottomrule
        \end{tabular}%
    }
\end{table}
\begin{table}[t]
    \centering
    \caption{Comparison of API costs for harness self-evolution across three datasets.}
    \label{tab:evolution-api-cost}
    \scalebox{0.85}{%
        \setlength{\tabcolsep}{8pt}%
        \begin{tabular}{c|l|c|c|c}
        \toprule
        \textbf{Dataset} &
        \textbf{Method} &
        \textbf{Training Evaluation} &
        \textbf{Evolution} &
        \textbf{Total} \\
        \midrule

        \multirow{3}{*}{\textbf{$\tau^2$-Airline}}
        & Self-Evolution
        & \$0.837
        & \$5.545
        & \textbf{\$6.382} \\
        & \textsc{Ecdysis} (w/o FDCR)
        & \$0.967
        & \$1.169
        & \textbf{\$2.136} \\
        & \textsc{Ecdysis} (w/ FDCR)
        & \$0.870
        & \$1.739
        & \textbf{\$2.609} \\

        \midrule

        \multirow{3}{*}{\textbf{$\tau^2$-Retail}}
        & Self-Evolution
        & \$0.609
        & \$7.875
        & \textbf{\$8.484} \\
        & \textsc{Ecdysis} (w/o FDCR)
        & \$0.555
        & \$1.930
        & \textbf{\$2.485} \\
        & \textsc{Ecdysis} (w/ FDCR)
        & \$0.506
        & \$5.257
        & \textbf{\$5.763} \\

        \midrule

        \multirow{3}{*}{\textbf{AgentBench}}
        & Self-Evolution
        & \$0.290
        & \$1.740
        & \textbf{\$2.030} \\
        & \textsc{Ecdysis} (w/o FDCR)
        & \$0.272
        & \$1.239
        & \textbf{\$1.511} \\
        & \textsc{Ecdysis} (w/ FDCR)
        & \$0.320
        & \$1.513
        & \textbf{\$1.833} \\

        \bottomrule
        \end{tabular}%
    }
\end{table}

\subsection{Model-Specific Accommodation (MSA)}

Table~\ref{tab:model-accommodation-analysis} summarizes the dataset-level and pooled MSA ratios.

\begin{table}[t]
\centering
\caption{
Model-accommodation ratios across datasets. The pooled ratio is calculated after merging all independent modification decisions, rather than by averaging the three dataset-level ratios.
}
\label{tab:model-accommodation-analysis}
\scalebox{0.95}{%
    \setlength{\tabcolsep}{8pt}%
    \begin{tabular}{l|c|c|c}
    \toprule
    \textbf{Dataset} &
    \textbf{Self-Evolution} &
    \textbf{\textsc{Ecdysis} (w/o FDCR)} &
    \textbf{\textsc{Ecdysis} (w/ FDCR)} \\
    \midrule
    $\tau^2$-Airline 
    & 67.65\% & 94.44\% & 40.00\% \\
    $\tau^2$-Retail
    & 43.75\% & 55.56\% & 50.00\% \\
    AgentBench    
    & 60.00\% & 80.00\% & 50.00\% \\
    \midrule
    \textbf{Overall}
    & \textbf{60.00\%}
    & \textbf{75.61\%}
    & \textbf{45.45\%} \\
    \bottomrule
    \end{tabular}%
}
\end{table}

\section{Analysis of Evolution Process}
\label{sec:evolution-process-analysis}

This section analyzes the actual execution of different harness evolution methods from four aspects: candidate generation, API cost, token usage, and end-to-end training time. All statistics are collected from the complete evolution runs described in Section~\ref{sec:experimental-setup}. The three methods use the same training evaluation, candidate validation, rollback, and early stopping protocols. We report the model calls and resource usage observed during actual execution. Therefore, these results reflect the cost of different candidate generation mechanisms along their actual execution paths rather than theoretical budgets normalized by a fixed number of evolution rounds.

\subsection{Candidate Generation}
\label{sec:candidate-generation}

Different evolution methods exhibit different candidate generation costs. Compared with SE, \textsc{Ecdysis} requires fewer coding-agent calls, mainly because the two methods use different update granularities. SE generates and accumulates modifications for individual failures, while \textsc{Ecdysis} aggregates the failure evidence within each evolution round and generates candidate modifications from the aggregated evidence. As the number of failures within a round increases, the candidate generation cost of SE grows accordingly, while \textsc{Ecdysis} continues to generate candidates at the round level. The runtime logs further reveal differences between the two methods during actual candidate generation. For $\tau^2$-Retail, approximately 67\% of the coding-agent calls from SE completed normally, compared with approximately 97\% in $\tau^2$-Airline. In contrast, all candidate generation calls from \textsc{Ecdysis} completed normally in both domains. These results show that accumulating modifications for individual failures not only increases candidate generation cost but is also associated with a higher proportion of incomplete calls in some runs. By aggregating failure evidence, \textsc{Ecdysis} reduces repeated candidate generation and exhibits more stable execution behavior in our experiments.

\subsection{API Cost and Token Usage}
\label{sec:api-cost-token-usage}

We divide the API cost of the complete training process into training evaluation and evolution (Fig.~\ref{fig:evolution-api-cost}). Training evaluation covers model calls made during training task execution. Evolution covers the cost of candidate analysis and implementation after failure evidence is generated, including calls to the coding agent, FDCR when enabled, and other model calls during the evolution process. Each cost is calculated from the input tokens, output tokens, and API prices applicable at execution time, as recorded in the call logs. The training evaluation cost remains below \$1 for all three methods, with most cost differences arising during evolution. SE incurs total costs of \$8.484 and \$6.382 on $\tau^2$-Retail and $\tau^2$-Airline, respectively. \textsc{Ecdysis} (w/o FDCR) reduces these costs to \$2.485 and \$2.136, corresponding to reductions of 70.71\% and 66.53\%. The total costs of \textsc{Ecdysis} (w/ FDCR) are \$5.763 and \$2.609, representing reductions of 32.07\% and 59.12\% relative to SE. Overall, both \textsc{Ecdysis} configurations incur lower total API costs than SE. This result is consistent with the reduction in coding-agent calls discussed above and suggests that generating candidates from failure evidence aggregated over an entire round can effectively reduce the cost of repeated implementation. SE achieves a pooled cache hit rate of 89.82\% across the three datasets (see Table~\ref{tab:pooled-evolution-cache-hit-rate}). In comparison, the pooled cache hit rate of \textsc{Ecdysis} (w/o FDCR) increases to 94.92\%, while \textsc{Ecdysis} (w/ FDCR) reaches 94.90\%. Notably, on $\tau^2$-Retail, \textsc{Ecdysis} (w/ FDCR) incurs substantially lower total API cost than SE despite using more input tokens. These results further demonstrate the cost efficiency of \textsc{Ecdysis}.

\subsection{End-to-End Evolution Time}
\label{sec:end-to-end-evolution-time}
We report the end-to-end training time from the initial training evaluation to the point when the final harness passes validation and is frozen in Table~\ref{tab:evolution-training-time}. For $\tau^2$-Retail, \textsc{Ecdysis} (w/o FDCR) and \textsc{Ecdysis} (w/ FDCR) achieve end-to-end training speedups of 1.42$\times$ and 1.30$\times$ over SE, respectively. For $\tau^2$-Airline, the corresponding speedups reach 3.23$\times$ and 1.84$\times$. These results indicate that generating candidates from failure evidence aggregated over an entire round reduces redundant work during evolution and consistently shortens end-to-end training time across both domains.


\end{document}